\documentclass[8.5pt,twoside,twocolumn]{article}
\usepackage[super,sort&compress,comma]{natbib}
\usepackage{mhchem}
\usepackage{times,mathptmx}
\usepackage{sectsty}
\usepackage{balance}
\usepackage[mathlines]{lineno}

\usepackage{graphicx} 
\usepackage{amsfonts}
\usepackage[psamsfonts]{amssymb}
\usepackage{lastpage}
\usepackage{url}
\usepackage[format=plain,justification=raggedright,singlelinecheck=false,
font=small,labelfont=bf,labelsep=space]{caption}
\usepackage{fancyhdr}
\usepackage[breaklinks=true,hyperindex=true,
colorlinks=true,pagebackref=false,citecolor=blue,plainpages=false,pdfpagelabels,
linkcolor=blue,urlcolor=blue]{hyperref}

\def\ip#1#2{\left\langle\,#1\, | \,#2\, \right\rangle}
\def\bra#1{\langle\,#1\,|}
\def\ket#1{|\,#1\, \rangle}

\def\e{{\mathrm e}}
\def\iu{{\mathrm i}}
\def\wn{~cm$^{-1}$}

\newcommand{\inp}[2]{\langle #1 | #2 \rangle }

\begin{document}

\thispagestyle{plain}
\fancypagestyle{plain}{
\fancyhead[L]{\includegraphics[height=8pt]{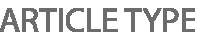}}
\fancyhead[C]{\hspace{-1cm}\includegraphics[height=20pt]{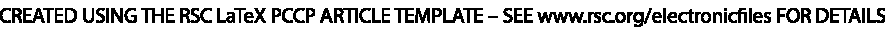}}
\fancyhead[R]{\includegraphics[height=10pt]{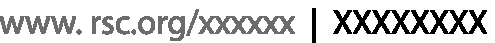}\vspace{-0.2cm}}
\renewcommand{\headrulewidth}{1pt}}
\renewcommand{\thefootnote}{\fnsymbol{footnote}}
\renewcommand\footnoterule{\vspace*{1pt}%
\hrule width 3.4in height 0.4pt \vspace*{5pt}}
\setcounter{secnumdepth}{5}

\makeatletter
\def\subsubsection{\@startsection{subsubsection}{3}{10pt}
{-1.25ex plus -1ex minus -.1ex}{0ex plus 0ex}{\normalsize\bf}}
\def\paragraph{\@startsection{paragraph}{4}{10pt}
{-1.25ex plus -1ex minus -.1ex}{0ex plus 0ex}{\normalsize\textit}}
\renewcommand\@biblabel[1]{#1}
\renewcommand\@makefntext[1]%
{\noindent\makebox[0pt][r]{\@thefnmark\,}#1}
\makeatother
\renewcommand{\figurename}{\small{Fig.}~}
\sectionfont{\large}
\subsectionfont{\normalsize}

\fancyfoot{}
\fancyfoot[LO,RE]{\vspace{-7pt}\includegraphics[height=9pt]{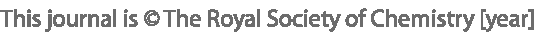}}
\fancyfoot[CO]{\vspace{-7.2pt}\hspace{12.2cm}\includegraphics{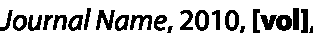}}
\fancyfoot[CE]{\vspace{-7.5pt}\hspace{-13.5cm}\includegraphics{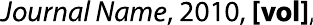}}
\fancyfoot[RO]{\footnotesize{\sffamily{1--\pageref{LastPage} ~\textbar  \hspace{2pt}\thepage}}}
\fancyfoot[LE]{\footnotesize{\sffamily{\thepage~\textbar\hspace{3.45cm} 1--\pageref{LastPage}}}}
\fancyhead{}
\renewcommand{\headrulewidth}{1pt}
\renewcommand{\footrulewidth}{1pt}
\setlength{\arrayrulewidth}{1pt}
\setlength{\columnsep}{6.5mm}
\setlength\bibsep{1pt}

\twocolumn[
  \begin{@twocolumnfalse}
\noindent\LARGE{\textbf{Computational Methodologies and Physical 
Insights into Electronic Energy Transfer in Photosynthetic 
Light-Harvesting Complexes
}}
\vspace{0.6cm}

\noindent\large{\textbf{Leonardo A. Pach\'on\textit{$^{a,b}$}
and
Paul Brumer$^{\ast}$\textit{$^{b}$}}}\vspace{0.5cm}

\noindent\textit{\small{\textbf{Received Xth March 2012,
Accepted Xth XXXXXXXXX 2012\newline
First published on the web Xth XXXXXXXXXX 2012}}}

\noindent \textbf{\small{DOI: 10.1039/b000000x}}
\vspace{0.6cm}

\noindent \normalsize{We examine computational techniques and methodologies
currently in use to explore electronic excitation
energy transfer in the context
of light-harvesting complexes in photosynthetic antenna systems, and comment
on some new insights into the underlying physics.
Advantages and pitfalls of these methodologies are discussed, as are
some physical insights into the photosynthetic dynamics.
By combining  results from molecular modelling of the complexes
(structural description) with an effective non-equilibrium statistical
description (time evolution), we  identify some general features,
regardless of the particular distribution in the protein scaffold, that are
central to light-harvesting dynamics and, that could ultimately be
related to the high efficiency of the overall process.
Based  on these general common features, some
possible new directions in the field are discussed.}

\vspace{0.5cm}
 \end{@twocolumnfalse}]



\begin{figure*}
\begin{minipage}[]{\columnwidth}
\includegraphics[width = \columnwidth]{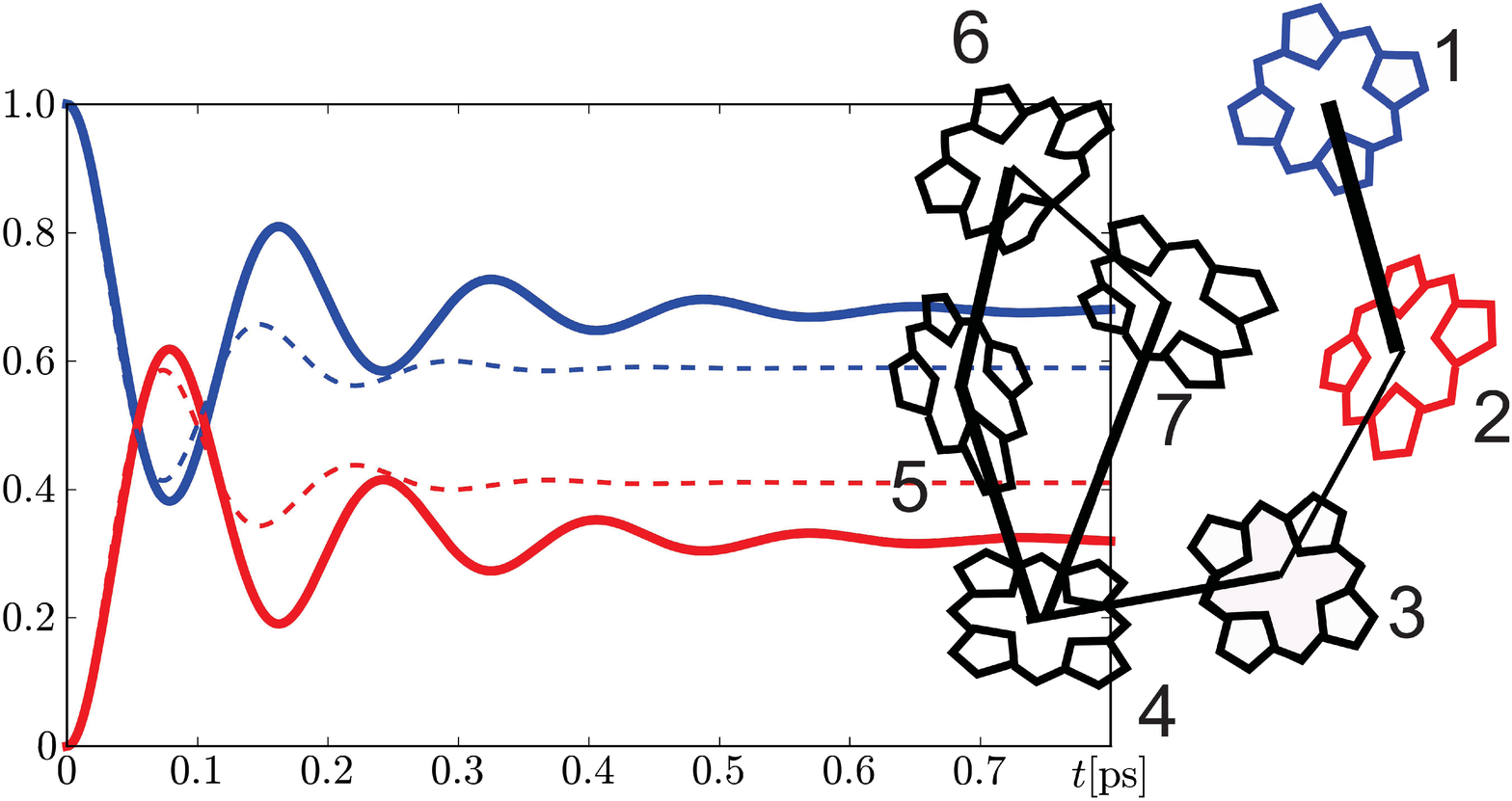} 
\end{minipage}
\begin{minipage}[]{\columnwidth}
\textbf{TOC entry:} Methodologies used to explore 
electronic energy transfer in light-harvesting 
photosynthetic complexes are examined and comments on new insights into the 
underlying physics are provided.
\end{minipage}
\end{figure*}

\section{Introduction}
The transfer of electronic excitation energy is ubiquitous in nature
and its dynamics and manipulation is of special interest in diverse
fields of physics, chemistry, biology and engineering.
Traditionally, electronic excitation energy transfer (EET) has been
studied and described in, e.g., molecular crystals \cite{JRS65}, rare
gases in the solid phase \cite{GJ&72,SKJ02} and dye aggregates
\cite{MD&95}.  More recently focus has turned to EET in biological
pigment-protein complexes present in light-harvesting complex of
natural photosynthetic antenna systems \cite{RMK01}.
The dynamics of the EET process in these systems is often thought of
in terms of donors and acceptors of excitation, and it is characterized,
mainly, by two time scales: a time scale $\tau_{\mathrm{da}}$ at
which the transfer between donors and acceptors takes place
(intermolecular transitions) and a second time scale $\tau_{\mathrm{vr}}$
at which the intramolecular (vibrational) relaxation occurs \cite{MK11}.

Based on the $\tau_{\mathrm{da}}$ and $\tau_{\mathrm{vr}}$ time
scales, the field distinguishes between two limiting cases. If the intermolecular
transitions are faster than the intramolecular relaxation,
$\tau_{\mathrm{vr}}$/$\tau_{\mathrm{da}}< 1$, then the exciton is
\emph{delocalized} among the donor-acceptor pairs and it travels
across the complex \emph{coherently}.
If, by contrast, $\tau_{\mathrm{vr}}$/$\tau_{\mathrm{da}}> 1$, then
the exciton is \emph{localized} and it hops from donors to acceptors,
and propagates \emph{incoherently} across the complex.
Under \textit{in-vivo} conditions, such as those in photosynthetic
light-harvesting complexes \cite{RMK01}, EET has been largely
considered as incoherent,
and its high efficiency ($\lesssim 100\%$) has
been discussed in terms of the ultrafast, tens of picoseconds, character
of the overall process \cite{SPv99,vVv00}.
[An important exception to the incoherent-dynamics assumption
is found in, e.g., the Light-Harvesting Complex II (LHCII is present in most of the 
higher plants) or in the Chlorosomes (found in, e.g., green sulfur bacteria) where, due to the strong 
coupling between chromophores, the excitations stay delocalized over a 
certain number of pigments, even at room temperature\cite{vVv00,SV&03}.]

Recent experimental results\cite{LCF07,EC&07,ME&09,CW&10,PH&10}
have suggested that during the first steps of EET in some light-harvesting
complexes, quantum coherences between different monomers of the
photosynthetic chromophores may  be assisting in the high efficiency of
the electronic excitation energy transfer.
This provocative suggestion resulted from the observation that these
coherences were, apparently, living longer ($\sim 800$~fs at 77~K and
$\sim400$~fs at 277~K) than could have been predicted (cf. 
Refs.~\citenum{CF09,PB11} or Sec.~\ref{sSec:WDoTLiveSoL} below).
These two issues, and their implications, rapidly became the
center of considerable study from diverse fields of chemistry
and physics (cf. the review articles in
Refs.~\citenum{Nv10,IC&10,OS11,MC11,SIF11,KN11}).

The initial attempts to explain the dynamics observed in these experiments
\cite{LCF07,EC&07,ME&09,CW&10,PH&10}
were based on F\"orster's theory \cite{Fos48,Fos65} (strong exciton-vibrational
coupling compared with the dipole-dipole interaction between the donor
and acceptor, fast intramolecular relaxation, incoherent transfer) or on the
Redfield formalism \cite{Red57} (weak exciton-vibrational coupling, slow
relaxation).
These studies revealed that these two approaches were inadequate 
because, in the dynamics of photosynthetic complexes, there
is neither a dominant time scale nor a leading coupling strength
\cite{Nv10,IC&10,OS11,MC11,SIF11,KN11}.  This fact motivated
the use of approaches and techniques developed in other fields such as
quantum optics \cite{OL&08,PH08,RM&09}, semiclassical molecular dynamics
\cite{HC10,TM10,HC11,HC11b}, solid state physics \cite{MW&11,PB11,NBT11},
etc., and encouraged the development of a variety of new techniques\cite{IC&10,SIF11}.
In addition, it motivated the improvement of molecular
modelling techniques \cite{Nv10,MC11,KN11}, in particular in aspects related
to the influence of the protein environment, and  
to the possibility of all-atom
simulations of the system time evolution \cite{SR&11}.

Our goal in this Perspective is to summarize the state-of-the-art
in the computational 
description of EET in photosynthetic complexes as well as to suggest
some new possible directions in the field.
In doing so, we collect  common features that have emerged
from the different  approaches.
We also attempt to collect the relevant physical information that
could be used as input for molecular modelling strategies
in the potential design of artificial EET systems.



\footnotetext{\textit{$^{a}$~Instituto de F\'isica,
Universidad de Antioquia, AA 1226, Medell\'in, Colombia.
Tel: 57 4 219 6439;
E-mail: leonardo.pachon@fisica.udea.edu.co}}
\footnotetext{\textit{$^{b}$~Chemical Physics Theory Group,
Department of Chemistry and
Center for Quantum Information and Quantum Control,
University of Toronto, Toronto, Canada M5S 3H6.
Fax: 1 416 978 5325; Tel: 1 416 978 7044;
E-mail: pbrumer@chem.utoronto.ca}}


\section{Photosynthetic Complexes}
Structures of natural 
light-harvesting antenna complexes are extremely
diverse.  
One can find anything from strcutures that are 
highly symmetric structures to  those
where the chromophores seem to be placed randomly in the protein
scaffold\cite{vVv00,Sch10,HC11}.
This diversity can be evidenced in, e.g., 
the interchromophore separation, which is, e.g.,
$\sim1$~nm for organisms with chlorophyll-based networks
or more than $\sim2$~nm in organisms with bilin-based 
networks\cite{vVv00,Sch10,HC11}.
Remarkably, in all cases the light harvesting efficiencies 
in these networks are reported to be close to 100\%\cite{vVv00,Sch10,HC11}.

Despite this variability in structure, the chromophore 
excitation energy/electronic
coupling landscape share some similarities that are worth analyzing 
in order to identify mechanisms that underlie the high efficiency of the
photosynthetic transport. 
We postpone this description until Sec.~\ref{sec:PhyIns} and, in the following, 
describe some of the structural features of two of the better studied complexes, 
the Fenna-Matthews-Olson (FMO) present in green sulfur bacteria and the 
Phycocyanin-645 (PC645) complex from \textit{Chroomonas} sp., 
in order to provide a physical picture of these systems.
Subsequently, we describe the standard theoretical framework for studying
EET\cite{MK11,RMK01}.

\subsection{Examples: FMO and PC645}
\label{sSec:MopDes}

The FMO pigment-protein complex from \emph{Chlorobium
tepidum} \cite{EC&07,PH&10} is a trimer consisting of
identical, weakly interacting monomers \cite{AR06}.
Each weakly interacting FMO monomer contains seven coupled
bacteriochlorophyll-$a$ (BChl-$a$) chromophores arranged asymmetrically,
yielding eight nondegenerate, delocalized molecular excited states (excitons)
\cite{EC&07,PH&10}. 
The coupling values of the seven chromophores, organized as 
$\{$BChl$a$~1, BChl$a$~2,..., BChl$a$~7$\}$,
can be summarized by the following Hamiltonian
\begin{footnotesize}
\begin{equation*}
H_{\mathrm{FMO}} = 
\left(
\begin{array}{ccccccc}
240 & -\mathbf{87.7} & 5.5 & -5.9 & 6.7 & -13.7 & -9.9 \\[0mm]
& 315 & \mathbf{30.8} & 8.2 & 0.7 & 11.8 & 4.3 \\[0mm]
& & 0 & -\mathbf{53.5} & -2.2 & -9.6 & 6.0 \\[0mm]
& & & 130 & -\mathbf{70.7} & -17.0 & -\mathbf{63.3} \\[0mm]
& & & & 285 & \mathbf{81.1} & -1.3 \\[0mm]
& & & & & 435 & 39.7 \\[0mm]
& & & & & & 245
\end{array}
\right),
\end{equation*}
\end{footnotesize}
where the energy of the third chromophore has be taken as the reference zero. 
Values are given in units of cm$^{-1}$ in the site representation\cite{AR06}.

Since the electronic coupling between the chromophores BChl$a$~1 and BChl$a$~2 is
relatively strong compared to the other coupling strengths \cite{AR06}, 
see $H_{\mathrm{FMO}}$,
it is common to focus, for purposes of model studies of EET, on the dynamics of the excitation in  a
dimer composed by BChl$a$~1 and BChl$a$~2. 
(This often adopted
model is discussed further in Sec.~\ref{sec:PhyIns}).

Although the seven site model has been the standard, 
an eight bacteriochlorophyll-$a$ 
chromophore was recently discovered\cite{TW&09,SM&11}.
In what follows we restrict our discussion to the case
of seven chromophores and refer the interested reader to Ref.~\citenum{MW&11} for a detailed
analysis of the influence of the eighth chromophore on the EET dynamics.

The PC645 pigment-protein complex has been studied both experimentally
\cite{CW&10} and numerically in great detail \cite{HC11}. It
contains eight bilin molecules covalently bound to the protein
scaffold. A dihydrobiliverdin (DBV) dimer is located at the center
of the complex and two mesobiliverdin (MBV) molecules located near
the protein periphery give rise to the upper half of the complex
absorption spectrum. Excitation of this dimer may initiate the light
harvesting process \cite{CW&10}. 
The coupling values of the eight bilin molecules, arranged as 
$\{$DBVc, DBVd, MBVa, MBVb, PCBc158, PCBd158, PCBc82, PCBs82$\}$,
can be summarized by the following Hamiltonian\cite{MD&07,CW&10}
\begin{footnotesize}
\begin{equation*}
\begin{split}
&H_{\mathrm{PC645}} = 
\\
&\left(
\begin{array}{lccccccc}
17034 & 319.4  & -9.6 & -43.9 & 20.3 & 25.3 & -46.8 & -20.0\\[0mm]
&  17116 & 43.9 & 7.7 & 30.5 & 29.0 & 21.5 & 48.0\\[0mm]
& & 16050 & 4.3 & -86.7 & -2.9 & -15.8 & 49.3\\[0mm]
& & & 16373 & 3.4 & 86.2 & 53.8 & -14.7\\[0mm]
& & & & 15808 & 7.8 & 11.0 & 10.0\\[0mm]
& & & & & 15889 & 29.0 & -10.7\\[0mm]
& & & & & & 15566 & 48.0\\[0mm]
& & & & & & & 15647
\end{array}
\right),
\end{split}
\end{equation*}
\end{footnotesize}
in units of cm$^{-1}$. 
The electronic coupling between
the closely spaced DBVc and DBVd molecules is $\sim$320 cm$^{-1}$.
This relatively strong coupling results in the delocalization of
the excitation, yielding the dimer electronic excited states labelled
DBV$+$ and DBV$-$. Excitation energy absorbed by the dimer flows
to the MBV molecules which are each 23 {\AA} from the closest DBV,
and ultimately to four phycocyanobilins (PCB) that absorb in the
lower-energy half of the absorption spectrum\cite{CW&10}.

Note  that  FMO  and  PC645  play  a distinctly different role
physically.  The  former  serves  to  transmit energy from the
antenna  to  the  reaction center.  That is, under natural conditions
light absorption is the function of the outer antenna  structure
and the probability of photon absorption  in the FMO is very low. 
By contrast, PC645 does, naturally,  play an 
active role in absorbing incident radiation.

\subsection{Hamiltonian}
\label{ssec:HowToDesDyn}
Theoretical modelling of the exciton energy transfer process
comprises intramolecular and intermolecular contributions (cf.
Chap~9 in Ref.~\citenum{MK11}, or Ref.~\citenum{RMK01}) and involves a number of assumptions, noted below, regarding the presumed physics.
For the particular case of a pigment-protein complex composed of $N$
pigments, the corresponding Hamiltonian, $H_{\mathrm{ppc}}$,
\begin{equation}
\label{EETGenHam}
H_{\mathrm{ppc}} = \sum_{m=1}^N H_{m}(\mathbf{r}_m,\mathbf{R}_m) +
\sum_{m,n\ge m}V_{mn}(\mathbf{r}_{mn},\mathbf{R}_{mn}),
\end{equation}
where $H_{m}$ describes  intra-pigment contributions and
$V_{mn}$ contains all the inter-pigment Coulomb interactions.
The Hamiltonian of the $m$-th pigment is characterized by the
electronic coordinates $\mathbf{r}_m$ and the nuclear coordinates
$\mathbf{R}_m$ and can be decomposed  as
$H_{m} = T_m^{(\mathrm{nuc})} + H_m^{(\mathrm{el})}$,
where $T_m^{(\mathrm{nuc})}$ denotes the nuclear kinetic energy
contribution.
Since we are not considering electron transfer among different
pigments, we can expand the pigment-protein complex Hamiltonian in
terms of the adiabatic electronic states
$\{\varphi_{ma}(\mathbf{r}_m;\mathbf{R}_m)\}$ defined by
$H_m^{(\mathrm{el})}(\mathbf{R}_m)
\varphi_{ma}(\mathbf{r}_m;\mathbf{R}_m) =
U_{ma} \varphi_{ma}(\mathbf{r}_m;\mathbf{R}_m)$, where $a$
refers to the $a$-th state of pigment $m$ and $U_{ma}(\mathbf{R}_m)$ denotes the
corresponding single-molecule potential energy surface (PEF).

The inter-pigment interactions in Eq.~(\ref{EETGenHam}) depend on
the distance $\mathbf{r}_{mn}$ between the electrons in the pigment
$m$ and those in the pigment $n$, as well as on distance $\mathbf{R}_{mn}$
between the electron in pigment $m$ and the nuclei of pigment $n$.
These interactions comprise the electron-electron interaction
$V_{mn}^{\mathrm{el-el}}$ and electron-nuclei interactions
$V_{mn}^{\mathrm{el-nuc}}$. As discussed in Sec.~\ref{ssSec:ElectDecoh},
the coupling to the nuclear degrees of freedom can be identified as the
first source of electronic decoherence.

The transfer of electronic excitation energy is often  accompanied by
the transfer of electrons\cite{Rat90,MK11}.
However, it is generally assumed that  the pigments are
sufficiently far apart so that the intermolecular
wave functions do not overlap 
$\left(\inp{\varphi_{ma}}{\varphi_{nb}} = \delta_{ma,nb}\right)$.
Under these circumstances, the
intermolecular interchange can be neglected. As such, we are neglecting
the bridge-mediated electron transfer (cf. Ref.~\citenum{Rat90} or Chap.~7
in Ref.~\citenum{MK11}) that could induce long-range super exchange of
electrons\cite{Rat90} . Nonetheless, one has to be aware that  strong
coupling to the solvent degrees of freedom can induce overlapping of the
intermolecular wave functions.

We shall also restrict ourselves to the situation when the electronic
ground state $S_0=g$ and the first excited singlet state $S_1=e$ of the
different pigments suffice for describing the dynamics of EET. 
This is the well known two-level model\cite{MK11}.
Additionally,  any non-adiabatic transitions between $S_0$
and $S_1$ during EET are neglected.

After projecting the Hamiltonian in Eq.~(\ref{EETGenHam}) onto the set
of states defined by the Hartree product ansatz
$\{\prod_{m=1}^N {\varphi_{ma_{m}}(\mathbf{r}_m;\mathbf{R}_m)}\}$, we
get\cite{MK11}
\begin{equation}
\label{equ:EETHamSimpV1}
\begin{split}
H_{\mathrm{ppc}} &= \sum_m \sum_{a=g,e} H_{ma}
\ket{\varphi_{ma}}\bra{\varphi_{ma}}
\\
&+ \sum_{m,n} J_{mn}
\ket{\varphi_{me}\varphi_{ng}}\bra{\varphi_{ne}\varphi_{mg}},
\end{split}
\end{equation}
where  the matrix elements
$H_{ma} = \bra{\varphi_{ma}}H_m\ket{\varphi_{ma}}$ and
$J_{mn} =
\bra{\varphi_{me}\varphi_{ng}}V_{mn}\ket{\varphi_{ne}\varphi_{mg}}$.

From the pigment-protein complex Hamiltonian in Eq.~(\ref{equ:EETHamSimpV1}),
it is clear that the electronic energies $H_{ma}$ as well as the excitonic
couplings $J_{mn}$ depend upon the nuclear coordinates.
As  discussed below, they also depend upon the external
environment (solvent), so that, in general,  their
values are constrained by the underlying interactions.  Methodologies and strategies for obtaining these parameters
are discussed in Sec.~\ref{Sec:MolMod}.

\subsubsection{The site representation}\textemdash
The transfer of electronic excitation energy can be described in
terms of the number of excitations present in the complex. If no
excitations are present, then the complex is in its ground state,
$\ket{0} = \prod_m \ket{\varphi_{mg}}$. In the case when a
single excitation resides at pigment $m$, 
it is denoted 
$\ket{m} = \ket{\varphi_{me}}\prod_{n\neq m} \ket{\varphi_{ng}}$.
This ordering of states suggests the following decomposition
of the Hamiltonian in Eq.~(\ref{equ:EETHamSimpV1}),
\begin{align}
H_{\mathrm{ppc}} &= H_{\mathrm{ppc}}^{(0)} + H_{\mathrm{ppc}}^{(1)} + \ldots  \nonumber
\\
\begin{split}
\label{equ:ExcHamExp}
&= \sum_{m}H_{mg}\ket{0}\bra{0}
\\&+
\sum_{m}\left(H_{me} + \sum_{n\neq m} H_{ng} \right)\ket{m}\bra{m}
+ \sum_{m,n}J_{mn}\ket{m}\bra{n}
\\&+ \ldots,
\end{split}
\end{align}
where the second line in Eq.~(\ref{equ:ExcHamExp}) denotes the Hamiltonian
of the zero-exciton manifold and the third line denotes the
single-exciton manifold Hamiltonian.
In Eq. (\ref{equ:ExcHamExp}), the intermolecular electrostatic coupling
has been neglected\cite{MK11}.
In general, the description of the dynamics in terms of the $n$-exciton manifolds 
is known as the site representation.

\subsubsection{Exciton-vibrational interactions}
\textemdash
As  already discussed,  in addition to the solvent, the vibrational degrees of freedom
comprise contributions from the single-pigment vibrations and from inter-pigment
vibrations\cite{MK11}.
The role of these vibrational degrees of freedom (DOF) is twofold: apart
from modulating the electronic energies and the excitonic couplings
in Eq.~(\ref{equ:EETHamSimpV1}) or in Eq.~(\ref{equ:ExcHamExp}),
they are a source of electronic decoherence in cases where the vibrations 
are not measured. As explained below, in such a case the electronic 
degrees of freedom constitute the system, whereas the vibrations serve
as the environment. Note that 
Sec. \ref{ssSec:ElectDecoh} contains a fulldiscussion of 
how the nuclear degrees of freedom
can induce electronic decoherence.

Note that, following Refs.~\citenum{RMK01} 
and \citenum{MK11}, we also effectively
include the effects of the vibrations as 
well as the influence of the solvent
(environment or reservoir) around the pigments.
In Sec.~\ref{sec:SouDissDec}, we discuss the characterization of this
effective model of the vibrations and the environment from a statistical
point of view.

\paragraph{Electronic decoherence by vibrational DOF.~~}
\label{ssSec:ElectDecoh}
As an example,  consider the excitation from an initial product state
comprising
the ground electronic state $\ket{g}$ and the ground vibrational state
$\ket{\nu_g}$ to an excited electronic state $\ket{e}$ in a single pigment.
For a sufficiently fast laser pulse, the state of the pigment will be described
by the density matrix
\begin{equation}
\begin{split}
\rho_{tot} &= A^2\left( \ket{g}\ket{\nu_g}\bra{\nu_g} \bra{g}
+ |c|^2 \ket{e} \ket{\nu_g(t)} \bra{\nu_g(t)} \bra{e}
\phantom{\e^{\iu \delta t\hbar}}\right.
\\ &\left.
+c^* \ket{g} \ket{\nu_g}\bra{\nu_g(t)} \bra{e}\e^{\iu \delta t\hbar}
+ c \ket{e} \ket{\nu_g(t)} \bra{\nu_g}\bra{g}\e^{-\iu \delta t\hbar}
\right),
\end{split}
\end{equation}
where $A$ is a normalization constant, $c$ is proportional
to the dipolar element $d_{e,g}$ and  $\delta$ is the
difference in energy between the minima in the nuclear potential
on the ground and excited electronic states.

Since our interest is in the electronic dynamics then, assuming the
Condon approximation is valid, we trace over the vibrational
states and get the density matrix for the electronic degrees of freedom:
\begin{equation}
\begin{split}
\rho &= A^2\left( \ket{g} \bra{g} + |c|^2 \ket{e} \bra{e}
+c^* \ket{g} \ip{\nu_g(t)}{\nu_g} \bra{e}\e^{\iu \delta t\hbar}
\right.
\\
&\left.
+ c \ket{e}\bra{g} \ip{\nu_g}{\nu_g(t)} \e^{-\iu \delta t\hbar}
\right).
\end{split}
\end{equation}
The electronic coherence, manifested in the off-diagonal matrix
elements $\ket{g}\bra{e}$ and $\ket{e}\bra{g}$, is modulated by
the overlap $\ip{\nu_g(t)}{\nu_g}$. As the vibrational wave packet
$\ket{\nu_g(t)}$ evolves on the upper electronic surface, 
particularly when the wavepacket comprises many states,  the overlap
$\ip{\nu_g(t)}{\nu_g}$ can decrease and so will the off-diagonal terms;
the result is electronic
decoherence (cf. Refs.~\citenum{PR97} and \citenum{PR98} or Chap.~6 in
Ref.~\citenum{SB11}).

This process, electronic decoherence by the vibrational states,
has been widely studied using quantum/classical methodologies
\cite{PR97,PR98,HR04b,FSB08,FB11} and it is expected to occur
on ultrashort time scales.
For example, results on betaine dye molecules \cite{HR04b} and
on femtosecond dynamics and laser control of charge transport in
trans-polyacetylene\cite{FSB08,FB11} suggest that these time scales
are very short, $\sim$2.5~fs and $\sim$3.7~fs, respectively.
We return to this subject in Sec.~\ref{sec:PhyIns}, because
it is the comparison of these time scales to those observed experimentally,
which are far longer, that has motivated a great deal of research.

In our description we have assumed that the each pigment is affected
only by its local vibrations, 
which are assumed to be uncorrelated. However, which degrees of freedom are to be regarded as the bath and what is the nature of the vibrations is crucial to the decoherence issue 
\cite{Sch07}.  For example, it has been shown that if the correlations between local vibrations
are taken into account, they could suppress the
cross-over to incoherent dynamics at high temperatures \cite{Naz09}
or affect the trapping time in rings of chromophores\cite{FNO10}.
This subject is widely discussed in Ref.~\citenum{NET10} and more
recently in Ref.~\citenum{SS11}, references to which the reader is referred.

\paragraph{Inclusion of the intra/inter vibrational DOFs
and the solvent.~~}\label{sssSec:EffHam}
As a first approximation, one could assume that
the nuclear motion as well as the low-frequency solvent coordinates
can be treated in the harmonic approximation and attempt to simulate
its effects via a set of dimensionless vibrational normal mode
coordinates $\{q_{\alpha}\}$ of frequency $\omega_\alpha$. 
In this case, the Hamiltonian for the ground state contribution of the 
pigment-protien-complex in Eq.~(\ref{equ:ExcHamExp}) now reads
\begin{align}
H_{\mathrm{ppc}}^{(0)} &=  \mathcal{H}_0\ket{0}\bra{0}
\\
\label{equ:calH0}
\mathcal{H}_0 &= E_0 + \sum_m H_{mg} 
+ \sum_{\alpha} \hbar \omega_{\alpha} b^{\dagger}_{\alpha}b_{\alpha}
\end{align}
where $E_0$ denotes the ground state energy, including vibrational 
zero-point energies, of the normal-mode vibrations\cite{MK11}.
$b^{\dagger}_{\alpha}$ and $b_{\alpha}$ are the annihilation and 
creation operators, respectively, expressed in terms of the normal mode
coordinates $Q_{\alpha} = (b^{\dagger}_{\alpha} + b_{\alpha})$
with $q_{\alpha} = \sqrt{\hbar/2 \omega_{\alpha}}Q_{\alpha}$.

Assuming that the ground and the singly excited electronic states can
be described by the same normal coordinates, the single excitonic
Hamiltonian reads\cite{MK11}
\begin{align}
\label{equ:H1ppc}
&H_{\mathrm{ppc}}^{(1)} = \sum_{m,n} \mathcal{H}_{mn} \ket{m}\bra{n}
\\
\label{equ:calHmn}
\begin{split}
&\mathcal{H}_{mn}= \delta_{mn} H_{me} + 
(1-\delta_{mn})\left[ J^{(0)}_{mn} +
\sum_{\alpha} \hbar \omega_{\alpha}\tilde{g}_{mn}(\alpha) Q_{\alpha} \right]
\\
&+\delta_{mn}
\left[
\sum_{k\neq m} H_{kg} + \sum_{\alpha} \hbar \omega_{\alpha}
\left(b^{\dagger}_{\alpha} b_{\alpha} +\frac{1}{2}+ g_m(\alpha) Q_{\alpha}\right)
\right]. 
\end{split}
\end{align}
The exciton-vibrational coupling matrix
$g_{mn}(\alpha) =
\delta_{m,n}g_{m}(\alpha) + (1-\delta_{m,n})\tilde{g}_{mn}(\alpha)$ in
Eq.~(\ref{equ:calHmn}) is given by the dimensionless coupling constants
$g_{m}(\alpha)$,  determined from the linearization of the potential
energy surfaces along the normal modes, and by the coupling constants
$\tilde{g}_{mn}(\alpha)$, which account for the influence of the vibrational
motions in the electronic couplings $J_{mn}^{0}$.
In the Frank-Condon approximation, vibrational motion does not induce
electronic transitions; that is,  $g_{mn}(\alpha) = \delta_{m,n}g_{m}(\alpha)$.

\subsection{Summary of the assumptions and approximations}
\label{sec:FreExcHam}
It is useful to summarize the approximation made in 
deriving Eqs.~(\ref{equ:calH0}) 
and (\ref{equ:calHmn}), which is essentially the
Frenkel-exciton model. Assumption one is the absence of inter-pigment 
electron transfer, which is guaranteed if there is no 
intermolecular wave function
overlap, e.g., if the pigments are sufficiently far apart. 
Second, the EET dynamics is assumed  to be 
fully characterized by the electronic ground state 
$S_0 = g$ and the first excited singlet state $S_1=e$ of each pigment,
yielding `` the two-level model".

In the next step one couples the electronic manifolds to the vibrational nuclear
modes and to the solvent coordinates, which induce
decoherence and dissipation. 
For simplicity, we have assumed no inter-pigment electronic-nuclear interactions,
so that nuclear vibrations pertain to each pigment. 
Additionally,  we assume that the collective nuclear-solvent vibrations 
can be described by a global manifold in its harmonic approximation.
Since the dipole-interaction matrix elements will depend, in general, upon this 
collective coordinate, we further adopt the Frank-Condon approximation
and neglect any possible non-adiabatic transitions induced by the solvent.
This set of approximation results in Eqs.~(\ref{equ:calH0}) 
and (\ref{equ:calHmn}).

\subsection{Dissipation and Decoherence}
\label{sec:SouDissDec}
Since the number of nuclear and solvent degrees of freedom (DOF)
in Eq.~(\ref{equ:H1ppc}) is extremely large, one could invoke a statistical
treatment of those DOF.  From a statistical  viewpoint the combined effects of the vibrational
states and the solvent in each pigment in Eq.~(\ref{equ:H1ppc}) can
be described in terms of a local fluctuating random force $\zeta_m(t)$
describing the fluctuations of the position-normal-mode coordinate.
Each of these fluctuating forces can be characterized in terms of
 its two-time correlation function
$\langle\zeta_m(t)\zeta_m(0)\rangle_{mg}$, where
$\langle \rangle_{mg}$ denotes statistical average over
the equilibrium density operator,
$\rho_{mg}^{\mathrm{eq}} = \mathrm{exp}(-H_{mg}\beta)/
\mathrm{tr}[\mathrm{exp}(-H_{mg}\beta)]$, of the
normal modes in Eq.~(\ref{equ:calHmn}).
For the particular case described in Eq.~(\ref{equ:calHmn}), the
correlation function is be defined by\cite{CL83,GSI88,Ing02,Wei08}
\begin{equation}
\label{equ:tpcf}
\begin{split}
\langle\zeta_m(t)\zeta_m(0)\rangle_{mg} &=
\hbar\int_0^{\infty}\frac{{\rm d} \omega}{\pi}J_m(\omega)
\\ & \left[ \coth\left(\frac{\hbar\beta\omega}{2}\right)\cos(\omega t)
-  {\rm i} \sin(\omega t)\right]\;.
\end{split}
\end{equation}
where $\beta = 1/ k_{\mathrm{B}} T$, $T$ is the temperature
of the combined environment and $J_m(\omega)$ is termed the
spectral density (see Sec.~\ref{sec:SpeDen} below).
The fact that $\langle\zeta_m(t)\zeta_m(0)\rangle_{mg}$ is complex
is a consequence of the anticommutativity of the normal mode coordinates
$p_{\alpha}$ and $q_{\alpha}$\cite{Ing02}. In the classical limit, $\hbar \rightarrow 0$,
the correlation function is real and becomes
\begin{equation}
\label{equ:dissker}
\langle\zeta_m(t)\zeta_m(0)\rangle_{mg} \rightarrow
\frac{2}{\pi\beta}\int_0^{\infty}\frac{{\rm d} \omega}{\omega}J_m(\omega)
\cos(\omega t) = \frac{1}{\beta}
\Gamma_{m}(t),
\end{equation}
where $\Gamma_m(t)$ is the relaxation function. $\Gamma_m(t)$ is
the directly observable quantity  in time-dependent fluorescence Stokes
shift experiments\cite{CF09,IC&10} and  is defined in terms of the
reorganization energy $\lambda_m$ as
$\Gamma_m(0) = 2\hbar \lambda_m$  \cite{Muk99}.
From $\Gamma_m(t)$, the relaxation
time $\tau_m$ of the environment can be determined as
$\tau_m=\frac{1}{\Gamma_m(0)}    \int_0^{\infty}   \mathrm{d}t
\Gamma_m(t)$.

\subsubsection{The spectral density.}\textemdash
\label{sec:SpeDen}
The spectral density contains the relevant information about the nature
of the environment and is usually chosen to fit some
measurable quantity, e.g., the optical spectrum, of the system.
As  discussed in Sec.~\ref{Sec:MolMod}, spectral densities may also
be obtained from computational methods (cf. Refs.~\citenum{AR06,OS&11a}).
Here again the idea is  that the dynamical features, usually classically described,
be in agreement with the optical measurable quantities.
The complexity of obtaining reliable spectral densities is indicated by the studies
in Ref.~\citenum{OS&11a}, where a host of highly structured spectral densities
are shown for FMO. In the following, we restrict attention to the
spectral densities already used for studying the dynamics of EET.

For the particular case of the FMO complex, e.g., Adolphs and
Renger\cite{AR06} derived a spectral density which allows calculations
that show agreement between the experimental optical spectrum and
calculations based on pigment transition energies.
This spectral density is given by
\begin{equation}
\label{equ:JwFMO}
J_{\mathrm{FMO}}(\omega) = \omega^2 S_0 g_0(\omega)
+ \omega^2 S_{\mathrm{H}} \delta(\omega-\omega_{\mathrm{H}}),
\end{equation}
with
$S_0 = 0.5$, $S_{\mathrm{H}} = 0.22$, $\omega_{\mathrm{H}}=180$\wn.
The spectral density $J_{\mathrm{FMO}}(\omega)$ comprises two parts,
one related to protein vibrations and characterized by
$
g_0(\omega) = 6.105\times 10^{-5}\times(\omega^3/{\omega_1}^4)
\e^{-\sqrt{\omega/\omega_1}}
+
3.8156\times 10^{-5}\times(\omega^3/{\omega_2}^4)
\e^{-\sqrt{\omega/\omega_2}}
$, $\omega_1 = 0.575$\wn  and $\omega_2=2$\wn, and the other
associated with  a vibrational mode characterized by the Dirac delta
function in Eq.~(\ref{equ:JwFMO}). This spectral function was
recently used in Ref.~\citenum{NBT11} to study the dynamics of the
FMO complex and the results are discussed in Sec.~\ref{ssSec:QUAPI}.
A comparison of this spectral density function and
that computed recently\cite{OS&11a} is provided in Sec. \ref{Sec:MolMod}
below.

Another type of spectral density widely used in the context of EET,
as well as in  non-linear spectroscopy, is the
Ohmic model with Lorentz-Drude regularization\cite{Muk99,GM05,Wei08,
CF09,IC&10,SIF11}
\begin{equation}
\label{equ:JwOhm}
J_{m}(\omega) =
 2\hbar \lambda_m \tau_m \omega
/\left(1 + \omega^2 \tau_m^2 \right)~.
\end{equation}
For this case, the relaxation function reads
$\Gamma_m(s) = 2 \hbar \lambda_m \exp(-s/\tau_m)$.
A variant to the Lorentz-Drude regularization model is the Ohmic spectral
density with exponential decay\cite{LC&87,Wei08,PB11}
\begin{equation}J_{m}(\omega)
=  2 K \omega \exp(-\omega/\omega_\mathrm{c}) , \label{ohmicJ} \end{equation}
where the dimensionless parameter $K$ describes the damping
strength and $\omega_{\mathrm{c}}$ is the cut-off frequency.
For this variant, the reorganization energy is given by
$\lambda_m = 2 K_m \hbar \omega_{\mathrm{c},m}$ and the
relaxation time by $\tau_m = \pi/(2\omega_{\mathrm{c},m})$.
The Ohmic spectral density with exponential decay is a member of a more
general family of spectral densities parametrized by $s$ as
$J(\omega) \sim \omega^s
\mathrm{e}^{-\omega/\omega_{\mathrm{c}}}$.
Here, $s=1$ is the Ohmic spectral density while $s<1$ ($s>1$)
is termed  the sub-Ohmic (super Ohmic) spectral densities\cite{Wei08}.
Although the Ohmic spectral density is a useful choice for, e.g., electron
transfer dynamics or biomolecular complexes \cite{MK11,GM06},
it has the drawback that it does not contain any high-frequency
modes of the environment.\cite{NM&08,NBT11,OS&11a}.

Another relevant spectral density is the one characterizing
the effect of blackbody radiation (e.g., natural sunlight or
moonlight)\cite{JB91,HB07,MV10,HB11,BS11,PB12}.
From an open-quantum-system perspective, the influence of the blackbody
radiation is condensed in the spectral density \cite{FLO85,FLO87,FLO88,BC91}
\begin{equation}
\label{equ:JwBB}
J_{\mathrm{BB}}(\omega) = M \tau_{\mathrm{BB}} \,
\omega^3 \Omega_{\mathrm{BB}}^2/
\left(\Omega_{\mathrm{BB}}^2 + \omega^2\right),
\end{equation}
where $M_{\mathrm{e}} =
m + M_{\mathrm{e}} \tau_{\mathrm{BB}} \Omega_{\mathrm{BB}}$ is the
renormalized mass of the electron (whose bare mass is $m$),
$\tau_{\mathrm{BB}} = 2 e^2 /3 M_\mathrm{e} c^3 \sim 6.24\times10^{-24}$s
and $\Omega_{\mathrm{BB}}$ is a frequency cutoff \cite{FLO85,FLO87,FLO88}.
For this spectral density, the decay function in Eq.~(\ref{equ:dissker})
is
$
\Gamma_{\mathrm{BB}}(s) =\tau_{\mathrm{BB}} \Omega_{\mathrm{BB}}^2
\left[2 \delta(s) - \Omega_{\mathrm{BB}} \exp(-\Omega_{\mathrm{BB}}|s|)\right].
$
In the limit $\Omega_{\mathrm{BB}} \rightarrow\infty$, one gets the
surprising result $\Gamma_{\mathrm{BB}}(s) = 0$. This  corresponds
to taking the point-electron limit.
An interesting feature of Eq. (\ref{equ:JwBB}) is that it is capable of introducing
fluctuations without dissipation, with no violation of the fluctuation-dissipation
theorem\cite{FO98}.
As such, it provides a 
sound statistical mechanical formulation for the pure-dephasing
approaches (cf. Refs.~\citenum{PH08,RM&09}).
We have explored the excitation of open quantum
systems by blackbody radiation in Ref.~\citenum{PB12} and some
associated comments are given in Sec.~\ref{sec:PhyIns}.

\section{Spectral-Density Based Approaches}
\label{sec:SpecDensApp}
The fact that in photosynthetic light-harvesting complexes there
is  neither a dominant time scale nor a leading coupling strength
\cite{Nv10,IC&10,OS11,MC11,SIF11,KN11} means that  modelling
the excitation energy transfer demands the use of robust techniques
beyond conventional theoretical approaches such as F\"orster's theory
\cite{Fos48,Fos65}, or Redfield\cite{Red57} or Lindblad equations\cite{Lin76,GF&78}.
Due to the  various approximations employed in these methodologies,
such as the use of perturbation theory in the system-environment coupling strength
\cite{CS05,IC&10,IF09a}, the incoherent
transfer approximation\cite{JNS04,IC&10,IF09a}, the assumption og Markovian dynamics
\cite{CS05,IF09a}, or the secular approximation\cite{IC&10,IF09a},
these approaches can provide inaccurate descriptions of  light-harvesting
dynamics.
They are therefore of limited applicability when treating
realistically parametrized models of light harvesting complexes\cite{HC10}.

Before discussing recent approaches, it is worth discussing
the most basic theory for EET: the F\"orster theory\cite{Fos48,Fos65}.
As already noted above, the theory developed by F\"orster applies
to the case of incoherent EET, i.e.,  to the case when
the intermolecular transfer times are slower than the intramolecular
relaxation.
This approach is based on the Fermi golden rule and on a second order
approximation to the excitonic coupling between pigments.
This implies that F\"orster's theory describes EET for weak exciton coupling.
F\"orster's theory characterizes EET in terms of excitation rates
$k_{m\rightarrow n}$ from pigment $m$ to $n$.
These rates are given in term of the spectral overlap between
the donor emission and acceptor absorption spectra and determined
by\cite{Fos48,Fos65,MK11}
\begin{equation}
\label{equ:FosRates}
k_{m\rightarrow n} = |J_{mn}|^2
\int\limits_{-\infty}^{\infty} \frac{\mathrm{d}\omega}{2\pi} \,
\Re[A_n(\omega )] \Re[F_m(\omega)],
\end{equation}
where
$F_m(\omega) =
\int_0^{\infty} \mathrm{d}t \,\mathrm{e}^{\iu \omega t}
\mathrm{e}^{-\iu(\omega_m - 2\lambda_m) t - g_m^*(t)}$
denotes the emission spectrum of the pigment $m$ and
$A_n(\omega)=
\int_0^{\infty} \mathrm{d}t \,\mathrm{e}^{\iu \omega t}
\mathrm{e}^{-\iu \omega_n t - g_n(t)}$
denotes the absorption spectrum of the pigment $n$. Here,
$g_m(t)$ is the line-broadening function\cite{Muk99,FC96,IC&10}.
Within the second-order cumulant expansikon (not required in 
the original Forster theory), $g_m(t)$ is given by:
\begin{equation}
\label{equ:gm}
g_m(t) = \frac{1}{\hbar^2}\int_0^t\mathrm{d}s_1 \int_0^{s_1}
\mathrm{d}s_2 \langle\zeta_m(s_2)\zeta_m(0)\rangle_{mg}~.
\end{equation}
This expression clearly shows that the dynamics, as well as the
transport properties, depend  on the statistics of the bath as well
as on the non-local (in time) character of the correlation
function.

We note that at present, in the related open-quantum-system community,
there is a great interest in quantifying the non-Markovian
(non-local in time) character of dynamics of given systems.
In doing so, some measures for the degree of non-Markovian behavior
have been given in Refs.~\citenum{BLP09} and \citenum{RHP10}.
An interesting connection betweemn our discussion above and 
that subject can be made as follows.
The intermediate integral over $s_2$ in Eq.~(\ref{equ:gm}) accounts
for the non-Markovian character of the dynamics.
This characteristic feature could be useful in order to \emph{experimentally}
quantify the non-Markovian character of a particular system based on its
line-broadening function. Work along this direction is currently in progress\cite{PB12b}.

Below, we divide the discussion into two main parts.
In Sec.~\ref{SubSec:MEBA}, we discuss some of the main methodologies
based on master equation descriptions while in Sec.~\ref{sec:PFBA}
we discussed those based on the propagating function.

\subsection{Master-equations}
\label{SubSec:MEBA}
Master equation approaches provide equations for the propagation of the density
matrix for the system dynamics (here the electrons),
obtained by tracing over the environmental degrees of freedom.
Popular master-equations based approaches for studying
EET are  the second-order perturbative time-convolution (TC2) and the second-order
perturbative-time convolutionless (TCL2) quantum master equations
(see Chap.~9 in Ref.~\citenum{BP02} for a formal basis  of these techniques)
and the hierarchical second-order cumulant expansion approach\cite{IF09,IF09a,IC&10}.
These approaches are based on the projection operator technique\cite{Nak58,Zwa60}.
The interested reader  may find useful the discussions in
Refs.~\citenum{Nak58,Zwa60,Zwa61,Gra82,BP02}.

A comprehensive and detailed discussion regarding the applicability of
these three methodologies using typical values present in photosynthetic
complexes, $J_{mn} \sim 100$~cm$^{-1}$,  $\lambda_{mn} \sim 100$~cm$^{-1}$,
$E_{m} \sim 100$~cm$^{-1}$ and $\tau_{m} \sim 50$~cm$^{-1}$
at 77 and 300 K for the dimer composed of $\ket{\varphi_{1e}}\ket{\varphi_{2g}}$
and  $\ket{\varphi_{1g}}\ket{\varphi_{2e}}$, was  presented in Ref.~\citenum{IC&10}.
For completeness, we briefly discuss only the main features of these
approaches.

-- TC2 is calculated in second-order perturbative approximation with respect
to the coupling of the system to the environment in the interaction picture\cite{Nak58,Zwa60,Gra82,BP02}.
Neglecting initial correlation between the pigments and the solvent/vibrational
DOFs, in this approach, the time evolution  of the system density matrix is calculated
as (cf. Chap.~9 in Ref.~\citenum{BP02})
\begin{equation}
\label{equ:NakZwaEqu}
\frac{\partial}{\partial t}\rho(t) =
\sum_{m=1}^N \int_{t_0}^t \mathrm{d}s \mathcal{K}_m(t,s) \rho(t),
\end{equation}
where $\mathcal{K}_m(t,s)$ is the convolution or memory kernel\cite{BP02}.
The equation (\ref{equ:NakZwaEqu}) is a simplified version of the
Nakajima-Zwanzig equation\cite{Nak58,Zwa60}. Although, the non-local
character of this integro-differential equation accounts for the non-Markovian
character of the dynamics, it also makes  Eq. (\ref{equ:NakZwaEqu})
cumbersome to implement. Thus, further approximations/assumptions
are needed.

Using a classical version of the fluctuation-dissipation theorem [equivalent
to taking the high temperature limit for the real part of the correlation function of the
noise in Eq.~(\ref{equ:tpcf})], it was shown in Ref.~\citenum{IC&10} that the
transition rates predicted by the TC2 equation deviate strongly from that
given by the Markovian Redfield equation as well as from those predicted
F\"orster theory\cite{IC&10}.
The authors in Ref.~\citenum{IC&10} conclude that the TC2 equation is
applicable only for the nearly Markovian regime, in spite of its non-Markovian
nature.

In Ref.~\citenum{SiB11} it was shown that the Markovian approximation to
the TC2 gives reliable results only for small reorganization energies
($\lambda_m< 10$~cm$^{-1}$ for typical values of EET in photosynthetic
complexes $\sim100$~cm$^{-1}$).  This is far from the domain relevant to
light harvesting.

-- TCL2 is based on the TC2 plus the application of the time-convolutionless
projector operator technique of Shibata \textit{et al.}\cite{SA80}, which allows
for a time local description of the master equation in Eq.~(\ref{equ:NakZwaEqu}).
TCL2 is accurate for describing coherence between electronic ground and excited
states in a monomer, regardless of the magnitude of the electron-phonon
coupling\cite{IC&10}.
However, as pointed out in Ref.~\citenum{IC&10}, TCL2  fails to describe
the transfer rate in a region of large reorganization energy,
$\lambda_m > 20$~cm$^{-1}$.
For small reorganization energies, the rate predicted by the TCL2 equation
is virtually the same as that for the Markovian Redfield equation\cite{IC&10}.

-- Second Order Cumulant and Hierarchy Expansion.
This strategy allows for the inclusion of site-dependent reorganization
energies as well as an appropriate description of the Gaussian fluctuations
of the bath\cite{IF09,IF09a,IC&10}.
On the basis of the high temperature approximation,
this approach has been used as a benchmark and as a reference in the
field because it is able to extrapolate between the Redfield and
F\"orster theories.
However, it is restricted to  bilinear system-bath coupling, and places very large
demands on memory\cite{HC11b}. Further, the numerical effort grows rapidly
with increasing system size, and with spectral densities that
deviate from the Lorentz-Drude form\cite{Str11}.
Notwithstanding, this approach has the advantage that can be implemented 
to be numerically exact and is currently used by many research groups (cf. Refs.~\citenum{IC&10,JS&11}).


Note that 77~K corresponds to $\sim 50$~cm$^{-1}$, and 300~K to
$\sim 200$~cm$^{-1}$. These are the typical temperatures at which the
2DPE experiments have been done.
One can immediately see that the high temperature
condition $\hbar E_m/k_{\mathrm{B}}T \ll 1$ adopted above in some approaches above
is barely fulfilled for light-harvesting
photosynthetic complexes ($\hbar E_m/k_{\mathrm{B}}T \sim$~2 at 77~K and
$\hbar E_m/k_{\mathrm{B}}T \sim$~0.5 at 300~K ).
In order to be applicable to other physical scenarios, where the deviations
from the high temperature are larger, it would be of interest to efficiently extend
these methodologies to the low temperature regime.

\subsection{Propagating-function methods}
\label{sec:PFBA}
The second main category of  approaches  relies
 directly on the propagating function of the system
and bath density matrix. We discuss below three of the most commonly
used: linearized-density-matrix-dynamics approaches,
the quasi-adiabatic propagator path integral (QUAPI) and the
non-interacting-blip approximation (NIBA).

\subsubsection{Linearized-density-matrix-dynamics}\label{ssSec:LinApp}
\textemdash
Linearized approaches \cite{MM79,ST97,TS99,BMC05,BC05,DBC08,HC10,
TM10,HC11,HC11b} for quantum evolution were developed in
order to deal with non-adiabatic chemical processes such as
proton and electron transfer in solution, excited-state molecular
fragmentation or molecular relaxation after electronic photo
excitation\cite{BMC05,BC05}. In general, these approaches assume
a quantum system represented by a number of discrete basis states
$\ket{n}$ coupled to an, in principle highly dimensional, environment
described by continuous coordinates $(\hat{Q},\hat{P})$ (cf.
Ref.~\citenum{HC10}).
The Hamiltonian for such a system, in the diabatic representation,
is given by
\begin{equation}
\label{equ:OriHalLinApp}
\begin{split}
 \hat{H} &= \hat{P}^2/2M + \sum_{\lambda}h_{\lambda\lambda}(\hat{Q})
\ket{\lambda}\bra{\lambda}
\\
& + \sum_{\lambda< \lambda^\prime}h_{\lambda\lambda^\prime}(\hat{Q})
\left( \ket{\lambda}\bra{\lambda^\prime} +
\ket{\lambda^\prime}\bra{\lambda} \right).
\end{split}
\end{equation}
The first term represents the nuclear kinetic term while the remaining
terms comprise the electronic Hamiltonian. Note that the Hamiltonian in
Eq.~(\ref{equ:OriHalLinApp}) is analogous to that in Eq.~(\ref{equ:H1ppc}).

An important ingredient in these approaches \cite{MM79,ST97,TS99,BMC05,
BC05,DBC08,HC10,TM10,HC11,HC11b} is to map the discrete
quantum states onto continuous coordinates. The
 strategy is based on  the mapping formalism\cite{MM79,ST97,TS99}. The
 idea is to replace the evolution of the electronic subsystem
with the evolution of a system of fictitious harmonic oscillators
\cite{BC05}. This is achieved in two steps: (i) Map the electronic
Hilbert space spanned by the $n$ diabatic states into one coinciding
with a subspace of $n$ harmonic oscillators of unit mass and
at most one quantum of excitation in a single specific oscillator
\cite{BC05}, i.e.,
$\ket{\alpha} \rightarrow \ket{m_\alpha} =
\ket{0_1,\ldots 1_\alpha \ldots,0_n}$,
and (ii) replace the projection operator by harmonic oscillator creation
and annihilation harmonic operators,
$\ket{\lambda}\bra{\lambda^\prime} \rightarrow
\hat{a}^{\dagger}_\lambda \hat{a}_{\lambda^\prime}$, and express
these in terms of their oscillator coordinates and momenta
$(\hat{q}_\alpha,\hat{p}_\alpha)$,
$ \hat{a}_{\lambda^\prime} = (
\hat{q}_{\lambda^\prime} + \iu \hat{p}_{\lambda^\prime}
)/\sqrt{2\hbar}$.
Once the mapping is performed, the original Hamiltonian in
Eq.~(\ref{equ:OriHalLinApp}) becomes
\begin{equation}
\label{equ:MapHalLinApp}
\begin{split}
\hat{H}_{\mathrm{m}} &= \hat{P}^2/2M
+ \frac{1}{2} \sum_{\lambda}h_{\lambda\lambda}(\hat{Q})
\left(\hat{q}_{\lambda}^2 + \hat{p}_{\lambda}^2 - \hbar\right)
\\
& +\frac{1}{2} \sum_{\lambda,\lambda^\prime}
h_{\lambda\lambda^\prime}(\hat{Q})
\left(\hat{q}_{\lambda}\hat{q}_{\lambda^\prime} +
\hat{p}_{\lambda}\hat{p}_{\lambda^\prime} \right).
\end{split}
\end{equation}
The density matrix is evolved according to
\newline
$\rho(t) = \e^{-\iu \hat{H}_{\mathrm{m}}  t/\hbar} \rho(0)
\e^{\iu \hat{H}_{\mathrm{m}}  t/\hbar}$. 
\newline
Thus, the resulting Hamiltonian
is expressed completely in terms of operators with continuous spectra.
Despite this fact, it is able to mimic the effects of transition among discrete
electronic states\cite{HC10,DBC08}.

The second main ingredient in the different linearized approaches is the
linearization itself. To do this, the unitary time-evolution operator
$\e^{-\iu \hat{H}_{\mathrm{m}}  t/\hbar} $ is represented in terms of discrete
phase space path integrals in the environmental variables and double sums over
the quantum states\cite{HC10,DBC08}. From the semiclassical perspective,
the evolution of the density matrix is  governed by forward and backward
trajectories associated with the unitary time-evolution operator and its adjoint
operator, respectively. The way in which the action is linearized in the different
approaches is what distinguishes between them; the particularities of each
 are discussed below.

In the linearized approximation to the semiclassical initial value representation
(LSC-IVR) of the unitary time-evolution operator (cf. Ref.~\citenum{TM10}),
the action in the path integral expression is expanded to linear order in the
distance between the forward and the backward trajectory of the environmental
and electronic degrees of freedom. Thus, the the effect of the bath (nuclear
degrees of freedom) on the dynamics of the system and the dynamics of the
systems itself are approximated. This approach was used
in the past\cite{MM79,ST97,TS99} and more recently used for the particular
case of the FMO complex by Tao and Miller\cite{TM10}. Despite the level of
approximation, their results for the dimer case are in agreement with
those of Ishizaki and Fleming\cite{IC&10,SIF11}. However, 
when the full complex is considered, the results at this level of 
approximation are not reliable \cite{HC11,HC11b}.

In the linearized approach to nonadiabatic dynamics using the mapping
formalism (LANDmap)\cite{BC05,HC10,HC11}, the linearization is performed
only in the environmental degrees of freedom. LANDmap has been used for studying
the dynamics of the FMO complex in Ref.~\citenum{HC10} as well as the
dynamics of the PC645 complex in Ref.~\citenum{HC11}. In particular, results
from Ref.~\citenum{HC10} show that when the bath relaxation is slow
(e.g., $\tau_{\mathrm{c}} \sim 500$~fs) results obtained with LANDmap are
reliable for times around 1~ps at high temperature ($\sim 300$~K) and for a
wide range of reorganization energies [$\lambda\sim0-5\Delta$,
being $\Delta$ the strength of the electric coupling ($\Delta=100$~cm$^{-1}$
in Ref.~\citenum{HC10})].
For fast relaxation (e.g. $\tau_{\mathrm{c}} \sim
100$~fs) and low temperature ($\sim 77$~K) LANDmap approach provides
a reasonable representation of the population oscillations only for short
times ($\sim100$~fs) and for small reorganization energies ($\lambda <
\Delta/5$); for this set of parameters, LANDmap is expected to fail to
reproduce thermalization\cite{HC10,HC11}.

LANDmap can be used to generate a short time propagator, which can be
iterated to generate more reliable results in the long time regime and valid
for a wider spectrum of parameters; this scheme is known as the iterative
linearized density matrix (ILDM) approach\cite{DBC08,HC10}. Although ILDM
is more accurate than LANDmap, its convergence after many iterations can be
problematic\cite{HC11b} and it is more demanding in computational terms.

The use of the mapping formalism brings its own drawbacks, in particular
when implemented and accompanied by semiclassical approximations
(cf. Ref.~\citenum{HC11b} for details).
To circumvent these problems, Huo and Coker have developed an improved
version of ILDM: the partial linearized density matrix (PLDM)\cite{HC11b}.
It is based on a coherent state representation of the electronic part and a
linearization of the nuclear (environmental) degrees of freedom.
This approach was tested for nonadiabatic multi states scattering problems
and it compared very well with standard results\cite{IC&10}.
An important feature of this new approach is the good performance at long
times and the correct equilibration with the thermal bath.
However, being based on a coherent state representation, PLDM suffers
from an  excess of free parameters\cite{DP09,DGP10,PID10}.  However,
the additional parameters can be used to improve the convergence of the 
average results.

The advantage of the linearized approaches is that they can be applied,
in principle, to any molecular model with arbitrary system bath interactions.
In particular, PLDM offers a favorable balance between accuracy and
computation efficiency\cite{HC11b}.

\subsubsection{QUAPI-based methods}
\label{ssSec:QUAPI}\textemdash
The quasi-adiabatic propagator path integral (QUAPI)\cite{MD95,MD95b,Tho00}
has been extensively used by Nalbach and Thorwart in the context of
EET\cite{TE&09,NI&11,NBT11}.
As in previous approaches, QUAPI assumes that the density matrix of the
 donor-acceptor system plus the environment can be characterized by a product
state at $t=0$, i.e. $\rho(0) = \rho_{\mathrm{da}}\otimes\rho_{\mathrm{env}}$.
The time evolution of $\rho_{\mathrm{da}}(t)$ is obtained after tracing out
the environmental (or bath) degrees of freedom, i.e.,
$
\rho_{\mathrm{da}}(t) \,=\, {\rm Tr} \left\{{U(t,0)\rho(0)U^{-1}(t,0)} \right\}_{\mathrm{B}},
$
$U(t,0)$ being the propagator of the full system plus bath.
The bath is modelled by a collection of harmonic modes
$H_{\mathrm{B}}=\frac{1}{2}\sum_{\ell} \left(p_{\ell}^2+\omega_\ell^2q_{\ell}^2 \right)$
in thermal equilibrium at temperature $T$.

QUAPI is based on a symmetric Trotter splitting of the short-time
propagator ${\cal K}(t_{k + 1}, t_k)$ for the full Hamiltonian into a
part depending on the system Hamiltonian and a part involving the
bath and the coupling term\cite{MD95,MD95b,NBT11}.
The splitting of propagator in terms of the short-time propagator is, by
construction, exact in the limit $\delta t = t_{k + 1} - t_k \to 0$.
However, for finite $\delta t$ it introduces a finite Trotter error, which has
to be eliminated by choosing $\delta t$ small enough until convergence is
achieved\cite{MD95,MD95b,NBT11}.

Another key ingredient in QUAPI's performance is the treatment of
non-Markovian time evolution generated from the non-local correlations
of the bath modes (see Sec.~\ref{sec:SouDissDec}).
For any finite temperature, these correlations decay exponentially fast
at asymptotic times, thereby setting the associated memory time
scale\cite{MD95,MD95b,NBT11}.
This fact allows for the introduction of an effective memory-time window
$\tau_{\rm mem} = \kappa \delta t$, so it is assumed that the system dynamics
have memory over $\kappa$ slices of time.
Over this time window, QUAPI defines an object called the reduced density
tensor, which has to be iterated in order to propagate the reduced density
matrix of the system.
Within the memory time window, all correlations are included exactly over
the finite memory time  $\tau_{\rm mem}$ and can, in principle, be neglected
for times beyond $\tau_{\rm mem}$.
Then, the memory parameter $\kappa$ has to be increased, until convergence
is found.
So, we have the requirement of decreasing $\delta t$ while increasing $\kappa$ in order
to get convergent results: the two strategies naturally run counter to one
another. However, convergent results can be obtained over a wide range of
parameter regimes\cite{MD95,MD95b,NBT11}.

The efficiency and implementation of the QUAPI algorithm are based
on the choice of the parameters $\kappa$ and $M$ (the number of basis
states), e.g., the reduced density tensor is a complex array of size
$M^{2\kappa+2}$. This limits its applicability to the case of
model systems that are not too large.
For the case of the donor-acceptor dimer system, $M=2$, so
with the standard hardware architectures one can choose $\kappa \lesssim
12 - 14$\cite{NET10,NI&11}.
However, if one is interested in studying, e.g., the network of eight 
bacteriochlorophyll-$a$ chromophores in the FMO complex within the two level system framework
one has $M=16$, so $\kappa \sim 3$. This relatively short memory window
could be problematic if one is interested in using this
approach in the context of electronic transfer, e.g., in semiconducting carbon
nanotubes\cite{HK&07}.

Although QUAPI is an iterative scheme, the truncation of the
memory makes the simulations  scale linearly with the propagation
time; however, since at low temperatures the bath induced correlations
 decay only algebraically\cite{GWT84,HR85,GSI88}, this truncation limits the applicability of QUAPI
to the finite temperature scenario\cite{MD95,MD95b,NBT11}.
In comparative studies\cite{NI&11,NBT11}, it has been shown that QUAPI
is able to generate the correct time evolution of photosynthetic complexes
in the range of parameters of interest, finite temperature, strong
system-bath interaction and strong electronic coupling.

QUAPI was implemented by Nalbach \emph{et al.} in Ref.~\citenum{NBT11}
in order to include the spectral density given in Eq.~(\ref{equ:JwFMO}) derived
in Ref.~\citenum{AR06}. The result suggests that the use of this spectral
density provides slightly smaller coherences lifetimes than those observed
in the experiment\cite{EC&07}, those calculated using the Ohmic spectral
density in Eq.~(\ref{equ:JwOhm}), and those based on a hybrid quantum/classical
all atom calculation\cite{SR&11}. It would be interesting to analyze this case
using other methodologies in order to understand the role of the vibrational
high frequency modes (see Sec.~\ref{sec:SpeDen}) in the dynamics.

\subsubsection{NIBA-based methods}
\label{ssSec:NIBA}\textemdash
In a series of papers, Cao \emph{et al.} \cite{WL&10,WL&11,MW&11} have
explored the efficiency of the energy transfer in light-harvesting systems using
approaches such the Haken-Strobl model (pure dephasing)\cite{HS73}, the
generalized Bloch-Redfield equation (second-order cumulant expansion model\cite{Cao97,WL&10})
and the non-interacting-blip approximation (NIBA)\cite{LC&87,Wei08}.
The Haken-Strobl model has been extensively discussed in the literature\cite{IC&10}
and the generalized Bloch-Redfield approach shares the spirit of the Nakajima-Zwanzig
equation discussed in Sec.~\ref{SubSec:MEBA}, so we discuss here the NIBA
approximation\cite{LC&87,Wei08} only.

To understand the term \emph{non-interacting blip}, we can appeal to the
path integral formulation of the systems with a finite Hilbert space\cite{LC&87,Wei08}.
For simplicity we restrict the discussion to the case of a system with two levels.
Time evolution of double-sided objects, like the density matrix, is performed by
a pair of trajectories\cite{Wei08,PID10}. In the case of the two level systems, these
trajectories are piecewise constant paths with sudden jumps between states, so
we have four possible ``positions'', namely both trajectories visit the ground or
excited state or one of them visits the ground state while the other one is the excited
state and vice-versa.
The two paths can be seen as a single path visiting the four configurations.
A period
of the path spent in the diagonal configuration (ground-ground or excited-excited) is
called a \emph{sojourn} while a period the path spends in an off-diagonal configuration
is a \emph{blip}\cite{LC&87,Wei08}.
The NIBA assumption is that the average time the system spends in a diagonal
configuration is much larger than time it spends in an off-diagonal one.
This assumption motivates setting (i) the sojourn-blip and (ii) the blip-blip correlations
to  zero; this leads to the non-interacting blip approximation.
In the strict Ohmic case, i.e., in the scaling limit $\omega_{\mathrm{c}} \rightarrow \infty$,
assumption (i) is \emph{exact}\cite{Wei08}; thus, the approximation enters, in this 
case, in neglecting the interblip interactions.

An alternate, apparently less complicated,
approach to deriving the NIBA is given in Ref. \citenum{Dek87}.

According to Ref.~\citenum{Wei08}, this approximation can be justified for: 
\begin{enumerate}
\item Weak-coupling to the environment and zero bias.
\item Super-Ohmic spectral densities with $s>1$ and $s>2$ at zero and finite temperature,
respectively.
\item Sub-Ohmic spectral densities with $s<1$ and $s<2$ at zero and finite
temperature, respectively.  Therefore, NIBA is justified\cite{LC&87,Wei08} (i) at all temperature 
in the sub-Ohmic case, (ii) at high temperature in the Ohmic case, and (iii)at high temperatures
in the super-Ohmic case with $s<2$.
 \cite{LC&87,Wei08}.
\end{enumerate}

NIBA can be formulated as a second-order master equation in the bath-dressed
electronic coupling\cite{APS86}, which is more convenient for the extension to
multilevel systems \cite{MW&11}.
This strategy was followed in Ref.~\citenum{MW&11} to generate a version of NIBA
for multi-state systems.
This extended NIBA was compared to the generalized Bloch-Redfield
approach by Cao \emph{et al.}\cite{MW&11} for describing the time evolution of
site populations in the FMO complex with eight bacteriochlorophyll-$a$ chromophores; 
the results are in excellent qualitative agreement.
NIBA's performance was also studied in Ref.~\citenum{ERT09} in a wide range of
parameters.  In the limits described above, it provides accurate results.

\section{Molecular Modelling of Photosynthetic Complexes}
\label{Sec:MolMod}
Molecular modelling of photosynthetic complexes is of primary importance
insofar as information about electronic couplings, site energies, spectral
densities, linear absorption spectrum, etc. are obtained from such studies\cite{AR06,
SC&07,MD&07,MC11,JZ&12}.
Recently, Mennucci and Curutchet\cite{MC11} presented a complete perspective
on these approaches and  K\"onig and Neugebauer\cite{KN11}
reviewed the advantages and pitfalls of most of these theoretical methods
in providing EET coupling constants.
Here we review  some of the most recent progress and physical insights
from the molecular-modelling studies.

As described in Ref.~\citenum{MC11}, among the various proposals
for molecular modelling one can distinguish two methodologies
for describing the environment and its effects on screening the Coulomb
interaction between the donor and acceptor transition dipoles.
The first method considers the environment as a dielectric continuum.
Its effects are characterized by means of a screening factor $1/n^2$, where
$n$ is the solvent refractive index (F\"orster-like theories).
The second approach treats the environment at the atomic level,
 either by using molecular mechanics force fields or by incorporating a full quantum
mechanical description of the chromophore-environment.

Mennucci {\it et al.} have developed a method,
reliant upon the dielectric-continuum-based approach, using a combination of the quantum linear response
and a structureless-polarizable-continuum-media model of the environment\cite{
TMC05,MC11}. The method is able to deal with the non-equilibrium response of the
system and of the environment during fast processes, such  as those involved
in electronic transitions and electronic energy transfer\cite{MC11}.
It has been successfully applied to examine the screening induced by the
environment in the electronic couplings for a set of over 100 chromophore
pairs, including chlorophylls, bilins and carotenoids, taken from structural
models of photosynthetic pigmentÐprotein complexes\cite{SC&07,MC11}.
An interesting outcome from this methodology is the fact that photosynthetic
light-harvesting can be tuned by heterogeneous polarizable environments
of the proteins\cite{CK&11} and that the final resonance energy transfer step
could occur on a timescale of 15 ps.
According to Ref.~\citenum{CK&11}, such a rapid final energy transfer step
cannot be reproduced by calculations based on the spectral density description,
which predict the energy transfer times to be on the order of 40 ps\cite{MD&07}.

The second method, where one carries out a
quantum all-atoms calculation	
for the structure and dynamics of photosynthetic complexes, is currently
beyond computational reach.
In order to overcome this problem, some methods based on quantum/classical
\cite{DK&02,CM&09,OK10,SR&11,OS&11} and continuum/atomistic descriptions
have been proposed\cite{MM&07,MC11}.
For a review of the main difficulties in the implementation of these approaches,
such as the inclusion of the polarizability of the solvent and the heterogeneous
character of the environment, the interested reader is referred to
Refs.~\citenum{MC11,KN11}.

Apart from the difficulties in computing the various electric couplings
using atomistic an description, a major challenge is the full-atom description
of the dynamics itself.
A first step towards an all-atom calculation of the dynamics of photosynthetic
complexes, in particular of the FMO complex, was presented by Aspuru-Guzik
\emph{et al.} in Ref.~\citenum{SR&11}.
Based on the Born-Oppenheimer approximation and adopting the Condon
approximation, they decomposed the total
system Hamiltonian into three parts: a system Hamiltonian acting on the
exciton sector (described by a set of two-level systems), a bath of
vibrational modes, and a coupling term between them; this corresponds
to the discussion in Sec.~\ref{sssSec:EffHam}.
The molecular energies are computed using time-dependent density functional
theory with the dynamics done by the Wigner method, i.e. the exciton dynamics
are done quantum mechanically and the bath dynamics are done
classically.
The evolution of the excitonic density matrix is obtained from a statistical
ensemble of unitary evolutions obtained by solving a time-dependent Schr\"odinger equation.
This approach allows consideration of nonidentical fluctuations across
all the sites.
The population dynamics of the chromophore is in accordance with
previous calculations\cite{IF09,IF09a,IC&10}.
Based on this approach, Aspuru-Guzik \emph{et al.}\cite{SR&11} concluded that the
site energy cross-correlation between chromophores does not play a significant
role in the energy transfer dynamics.
This is in accord with the  results of Olbrich in
Refs.~\citenum{OK10,OS&11} for the FMO complex and with those of Huo
and Coker\cite{HC11} for the PC645 complex.

It is worth mentioning that the statistical-ensemble-averaging approach has
inspired the application of quantum state diffusion approaches such as those
discussed in Refs.~\citenum{YDG&99,SG02}. In particular, it has been used to study the
role of quantum oscillations as well as the dependence on site energies
in electronic excitation transfer in the FMO complex\cite{RR&11},
the influence of the vibrational modes\cite{RR&11b} and the influence of noise,
disorder, and temperature on localization in excitonic systems\cite{MZC12}.
These techniques appear accurate,  efficient and are valid
for a wide spectrum of parameters\cite{RR&11,RR&11b,MZC12}.

In addition, we note that direct connections are being established between
all-atoms calculations and master equation approaches that use
spectral densities\cite{OS&10,OS&11a}. For example, Ref. \citenum{OS&11a}
obtained spectral densities associated with FMO by such a full-atom
computation. Results, whose effect on FMO dynamics is yet to be determined,
show structured $J(\omega)$ with considerably higher values of $\omega$
contributing to the spectral density than assumed in the Ohmic or
Drude-Lorentz models or Adolphs-Renger model\cite{AR06} previously utilized.

Finally, note that recent evidence has been found for correlated fluctuations of site energies and intersite electronic couplings as well as electronic-electronic coupling that could be more significant than the apparently uncorrelated site energy fluctuations \cite{OS&11,HC12}  Hence, an interesting possible direction in this regard is to develop technique which provide spectral densities that include bath-dependent electronic terms \cite{HC12}.

\section{Insights into the Observed Coherences}
\label{sec:PhyIns}
Below we discuss some of the
issues relating to the role of the observed EET  dynamics in natural
photosynthetic processes, and
introduce an analytic  model that provides physical insight into the origins of
the observed long-lived coherences.

\subsection{Natural processes}
We remark on two issues in need of considerable clarification if we are
to understand the role of the observed coherences in nature.
First, note that under natural conditions, photosynthetic complexes are excited by
sunlight, an incoherent source of light\cite{JB91,HB07,MV10,HB11,BS11,PB12}
that is incident on the system for times that are huge compared to the
time scale of the molecular dynamics.
As a consequence, excitation of this kind cannot generate coherent dynamics
among the pigments, neither in the case of the unitary dynamics associated with
isolated molecules, \cite{JB91,HB07,HB11,BS11} nor in the case of open systems,
where the molecule is in contact with an environment\cite{MV10,PB12}.
Based on this argument, it is not yet clear whether coherences
induced by coherent femtosecond laser pulses can play
a role under \textit{in vivo} conditions.

Second, there remains a concern that the observed coherences may not necessarily
be solely electronic in nature, i.e., suggestions have been made that vibrational
coherences are being observed as well
(cf. Ref.~\citenum{CK&12}).
Only very recently has an experimental protocol been
applied that, based on the nature of the observed 2D Photon Echo experiment,
can distinguish between electronic and vibrational coherences.
A recent application of  this protocol\cite{TD&12} for the case of PC645 has identified
one of the observed long-lived coherences as electronic, and the others as
likely vibrational. In this regard we note that vibrational
coherences would be expected to decohere slowly\cite{EB04} so that
understanding such a coherence feature presents no significant qualitative
challenge.

\subsection{The origin of long-lived coherences}
\label{sSec:WDoTLiveSoL}

The major part of this paper has provided a review of
a number of techniques and approaches for the study of
the EET dynamics of photosynthetic complexes.
Examination of these detailed computations shows, however,  that they
do not provide direct insight into the basic physical
origin of the observed longevity of the   coherences.
Below we address the question as to
the physical origin of the long-lived observed coherences, by introducing an
analytically soluble model that provides correct results for dimers that
characterize the FMO and PC645 systems.

At high temperatures and for weak coupling to the environment, a
classical treatment of the thermal fluctuations with an Ohmic spectral
density with exponential decay (see Sec.~\ref{sec:SpeDen}) predicts
that the electronic coherence discussed in Sec.~\ref{ssSec:ElectDecoh}
decays at a \emph{Gaussian} rate that can be determined as\cite{HR04,GM08,CF09}
\begin{equation}
\label{equ:tGrate}
\tau_{\mathrm{G}} = \sqrt{\hbar^2/2\lambda k_{\mathrm{B}} T},
\end{equation}
where $\lambda$ is the system reorganization energy.
Based on this expression, the dephasing time for photosynthetic
complexes, for a typical value of $\lambda = 130$~cm$^{-1}$),
can be estimated \cite{CF09} to be $\tau_{\mathrm{G}} = 45$~fs at
$T=77$~K and $\tau_{\mathrm{G}} = 23$~fs and $T=294$~K.

By contrast, the experiments in photosynthetic complexes
such as the FMO complex\cite{EC&07,PH&10} and the PC645
complex\cite{CW&10}, discussed above, have found that electronic
coherences among different chromophores survive up to 800~fs at
$77$~K \cite{EC&07} and up 400~fs at room temperature
\cite{CW&10}. Generally longer coherence times have been observed for
FMO in Ref.~\citenum{PH&10}.
This surprising observation and its possible consequences for
biological processes have been discussed extensively
\cite{EC&07,PH08,PH&10,CF09,RM&09,IF09,WL&10,CW&10,BE11,
HC11,SR&11,NBT11} and have provided motivation for the
development of the methodologies discussed above.
Interestingly, despite the diversity of approaches
and techniques, most \cite{CF09,RM&09,IF09,CW&10,BE11,HC11,
SR&11,NBT11,PB11} predict long-lived coherences on the same
time scales as those found experimentally \cite{EC&07,CW&10,PH&10}.
This suggests that the underlying physical features are correctly contained
in these approaches.
However, the sheer complexity of these computations has limited
one from identifying these essential physical features\cite{PB11}.

In order to explore the physical features responsible for the
survival of these coherences, we (in
Ref.~\citenum{PB11}) discuss the case of the relatively strongly coupled dimer composed of the
BChl1$a$ 1 and BChl1$a$ 2 in the FMO complex (see Sec.~\ref{sSec:MopDes}),
and the dimer formed of chromophores DVBc and DVBd in the PC645
complex (see also Sec.~\ref{sSec:MopDes}). The generic Hamiltonian for
the two two-level system is described in terms of Pauli spin matrices by\cite{GM05,GM06}
\begin{align}
\label{equ:DefInitHamilt}
\begin{split}
H &= \frac{\hbar}{2}\epsilon_1 \sigma_{z,1} + \frac{\hbar}{2}\epsilon_2 \sigma_{z,2}
  + \frac{\hbar}{2}\Delta\left(\sigma_{x,1} \sigma_{x,2} + \sigma_{y,1} \sigma_{y,2} \right)
\\
  &+\frac{\hbar}{2}\delta\mu_1\sigma_{z,1} R_1 + \frac{\hbar}{2}
\delta\mu_2\sigma_{z,2} R_2
  + B_1 + B_2,
\end{split}
\end{align}
where
$R_{i} = \sum_{\alpha}C_{\alpha,i}\left(a_{\alpha,i}
+ a_{\alpha,i}^{\dagger}\right)$
is the reaction field operator for molecule $i$,
$B_{i} = \sum_{\alpha}\hbar \omega_{\alpha,i}
a_{\alpha,i}^{\dagger} a_{\alpha,i}$
is the energy stored in the solvent cage of molecule $i$ and $\delta\mu_j$
is the difference between the dipole moment of the chromophore $j$
in the ground and excited states \cite{GM05,GM06}.
The first two terms in Eq.~(\ref{equ:DefInitHamilt}) are the
contributions from the individual sites and the third term is the $\Delta$
coupling between them.
The subsequent terms describe the system-bath coupling.
Following Refs.~\citenum{GM06} and \citenum{ERT09}, the Hamiltonian in
Eq.~(\ref{equ:DefInitHamilt}) can be written with respect
to the basis
$\{|g_1\rangle \otimes |g_2\rangle, |g_1\rangle \otimes
|e_2\rangle, |e_1\rangle \otimes |g_2\rangle,|e_1\rangle \otimes
|e_2\rangle \}$ describing the state of the two chromophores, i.e.

\begin{equation}
\label{eq:H_2C}
\begin{split}
&H=\sum_{i=1,2}\sum_{\alpha}
\hbar\omega_{\alpha,i}a_{\alpha,i}^{\dagger}a_{\alpha,i}
 \\
&+\frac{\hbar}{2}\left(
\begin{array}{cccc}
-\left(\epsilon_+ + V_+\right) & 0 & 0& 0   \\
0& -\left(\epsilon_- + V_-\right) & 2\Delta & 0\\
0 & 2\Delta & \epsilon_- + V_- &0\\ \nonumber
0 & 0 & 0 & \epsilon_+ + V_+
\end{array}\right)\,,
\end{split}
\end{equation}

where $\epsilon_{\pm}\equiv\epsilon_1\pm\epsilon_2$, and
$V_{\pm}\equiv \delta\mu_1 R_1 \pm \delta\mu_2  R_2 $.

\begin{figure}
\begin{center}
\includegraphics[width = 0.8\columnwidth]{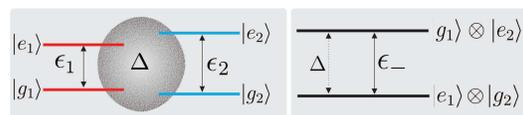}
\caption{Left hand side: The pair of interacting chromophores.
Right hand side: the
effective light harvesting two-level system
formed from the pair of
interacting chromophores \cite{GM06,ERT09}.}
\label{qubit}
\end{center}
\end{figure}

Since under excitation by weak light only the singly excited
states need to be taken into account, we can identify\cite{GM06,ERT09}
the active environment coupled 2D-subspace as
$\{|e_1\rangle\otimes|g_2\rangle,|g_1\rangle\otimes|e_2\rangle\}$.
In this central subspace of Eq.~(\ref{eq:H_2C}), the effective
interacting biomolecular two-level system Hamiltonian reads
\begin{equation}
\label{eq:Baqubit}
H=\left(\frac{\hbar\epsilon}{2}\sigma_z+\hbar\Delta\sigma_x\right)+
\frac{\hbar}{2}\sigma_z V +
\sum_{\stackrel{\alpha}{i=1,2}}
\hbar\omega_{\alpha,i}a_{\alpha,i}^{\dagger}a_{\alpha,i}\,
\end{equation}
where $\epsilon \equiv \epsilon_-$ and $V \equiv V_-$.
This is schematically illustrated in Fig.~\ref{qubit}, where $\Delta$ is
the associated ``tunneling energy", between the new basis states
$|e_1\rangle\otimes|g_2\rangle$ and $|g_1\rangle\otimes|e_2\rangle$.

Following the description given in Sec.~\ref{ssec:HowToDesDyn}, we assume
that the two baths are uncorrelated
$\langle R_1(t'')R_2(t')\rangle = \langle R_2(t'')R_1(t')\rangle = 0$.
Hence, Eq,~(\ref{eq:Baqubit}) can be written in the standard form of
the spin-boson model \cite{GM06}
\begin{equation}
\label{eq:Bqubit}
H=\left(\frac{\hbar\epsilon}{2}\sigma_z+\hbar\Delta\sigma_x\right)+
\frac{\hbar}{2}\sigma_z \sum_{\beta}g_{\beta}(b_{\beta} + b_{\beta}^{\dagger}) +
\sum_{\beta}\hbar\omega_{\beta}b_{\beta}^{\dagger}b_{\beta}\ ,
\end{equation}
where the $b_{\beta}$ includes harmonic oscillators coupled to both
chromophores, with couplings $g_{\beta}$. The effects of the baths are
treated in terms of the Ohmic spectral density with exponential decay
presented in Sec.~\ref{sec:SpeDen}.

The Hamiltonian in Eq.~(\ref{eq:Bqubit}) differs from the one described
in Sec.~\ref{sssSec:EffHam} by the fact that in Eq.~(\ref{eq:Bqubit}),
both states $\ket{e_1}\otimes\ket{g_2}$ and $\ket{g_1}\otimes\ket{e_2}$
are coupled to the same effective bath modes, whereas the treatment 
presented in Sec.~\ref{sssSec:EffHam} shows that for the case of a pair
of donor-acceptor pigments, the ground and excited states should be coupled
to independent modes. 
Although, this fact introduces fluctuations of the reference energy of the donor-acceptor 
system, if one is interested only in the dynamics of the Bloch oscillation this is irrelevant 
(cf. Ref.~\citenum{GM05,NET10}).
For strong coupling to the bath or high degree of non-Markovian dynamics, one expects
the two models, e.g, are driven to different thermal states.
However, the fact that our results, see below, coincide with the overwhelming majority 
of the current results in literature, means that the coupling strength and the degree of 
non-Markovian character, for the particular cases discussed below, are still in a regime 
where both models coincide. 
Note also that the previous predictions made for the 
lifetimes (cf. Ref.~\citenum{CF09}) were based on the spin boson-model. 
Thus, in order to get an idea about the lifetimes of the coherences, the model is 
quite well justified.

The parameter range within which the dimers described above, BChl1$a$ 1 and 
BChl1$a$ 2 in the FMO complex and the chromophores DVBc and DVBd in the 
PC645 complex, lie allows for the use of the non-Markovian NIBA plus first order 
corrections  in the interblip correlation strength, i.e. an {\it enhanced NIBA approximation}. 
For a description of the range of applicability of the \emph{bare} NIBA cf. 
Sec.~\ref{ssSec:NIBA}.
The enhanced approximation is valid for weak system-bath coupling
and for $\epsilon/2\Delta < 1$, over the whole range of temperatures (see
Chap.~21 in Ref.~\citenum{Wei08}), and provides simple and accurate
\emph{analytic} expressions for relaxation and decoherence rates.
\begin{table}[h!]
\begin{center}
\begin{tabular}{c|c|c|c|c|c}\hline
$K$ & $\epsilon/2\Delta$ &  $k_{\mathrm{B}} T/ 2 \hbar \Delta$ &
 $2\Delta/\omega_{\mathrm{c}}$ & $k_{\mathrm{B}} T / \hbar \omega_{\mathrm{c}}$
 & $\epsilon/ \omega_{\mathrm{c}}$
\\
\hline \hline 0.105 & 0.428 &  0.305 & 1.052 & 0.321 & 0.45\\ \hline
0.105 &  0.428 & 1.098 & 1.052 & 1.154 & 0.45\\ \hline
\end{tabular}
\end{center}
\caption{Parameters used for dimer formed of BChl$a$~1 and
BChl$a$~2 at $T=77$~K (first row) and
$T=277$~K (second row).}
\label{tab:parametersFMO}
\end{table}
As it was discussed in Sec.~\ref{ssSec:NIBA}, in the limit when $\omega_{\mathrm{c}}$ 
is much larger than the other frequencies in the system, NIBA is exact. 
In the present case, see Table~\ref{tab:parametersFMO}, this condition is not
fulfilled in all cases; however, we do have here
an improved approximation which provides lifetime estimates that are in very 
good agreement with experimental as well more-refined-theoretical results.

The high temperature limit in this approach is given by
temperatures well in excess of
$T_{\mathrm{b}} = \hbar (\Delta_{\mathrm{eff}}^2 + \epsilon^2)^{1/2}/k_{\mathrm{B}}$, where
\begin{equation*}
\Delta_{\mathrm{eff}} =
[\Gamma(1-2K)\cos(\pi K)]^{1/2(1-K)}(\tilde{\Delta}/\omega_{\mathrm{c}})^{K/(1-K)}\tilde{\Delta},
\end{equation*}
where in our case $\tilde{\Delta} = 2 \Delta$. Here the spectral density is ohmic [Eq. (\ref{ohmicJ})] 
where  $K$ describes the coupling strength to the bath and  $\omega_c$ is the 
bath cutoff frequency.

For the set of parameters listed in Table~\ref{tab:parametersFMO},
we find $T_{\mathrm{b}} \approx 288$~K.
Hence the FMO experiments, at 77 K and 277 K, are in the low
temperature regime, $T < T_{\mathrm{b}}$.
In this regime, the Rabi frequency $\Omega$, the relaxation rate
$\gamma_{\mathrm{r}}$ and the decoherence rate $\gamma$
are given by \cite{Wei08}
\begin{align}
\begin{split}
\label{equ:Omega}
\Omega^2 &=
\Delta_{\mathrm{b}}^2
\\ & + 2 K \Delta_{\mathrm{eff}}^2
           \left[\Re\psi(\mathrm{i}\hbar\Delta_{\mathrm{b}}/2\pi k_{\mathrm{B}} T)
     - \ln(\hbar\Delta_{\mathrm{b}}/2\pi k_{\mathrm{B}} T)\right]
\end{split}
\\
\label{equ:gammar}
\gamma_{\mathrm{r}} &=
\pi K \coth(\hbar \Delta_{\mathrm{b}}/2k_{\mathrm{B}}T)
\Delta_{\mathrm{eff}}^2/\Delta_{\mathrm{b}},
\\
\label{equ:gamma}
\gamma &=
\gamma_{\mathrm{r}}/2
+ 2 \pi K (\epsilon^2/\Delta_{\mathrm{b}}^2)k_{\mathrm{B}} T/\hbar,
\end{align}
respectively, where
$\Delta_{\mathrm{b}} = \sqrt{\Delta_{\mathrm{eff}}^2 + \epsilon^2}$ and
$\psi(z)$ is the digamma function.
(Note that in Refs.~\citenum{WW89} and \citenum{Wei08}, the term $\Delta_{\mathrm{eff}}^2$
is missing in the expression for $\Omega$.)
The expressions in Eqs. (\ref{equ:Omega})-(\ref{equ:gamma})
were derived for the particular case of non-Markovian Ohmic dissipation,
however, analogous expressions for arbitrary spectral densities can be
found in  Refs.~\citenum{WW89,Wei08,PB11}.

Based on this model, and using the parameters in Table~\ref{tab:parametersFMO}
(obtained from experimental results\cite{PB11}),
we find \cite{PB11} the following time scales
for the FMO complex
$2 \pi \Omega^{-1} = 163$~fs, $\gamma_{\mathrm{r}}^{-1} = 90$~fs, $\gamma^{-1} =
153$~fs at $T=77$~K, while
$2 \pi \Omega^{-1} = 151$~fs, $\gamma_{\mathrm{r}}^{-1} = 45$~fs,
$\gamma^{-1} = 69$~fs at $T=277$~K.
For the case the dimer formed of chromophores DVBc and DVBd in the
PC645 complex, we predicted $2 \pi \Omega^{-1} = 49$~fs,
$\gamma_{\mathrm{r}}^{-1} = 76$~fs and $\gamma^{-1} = 88$~fs
at $T=294$~K using the parameters in Refs.~\citenum{MD&07,HC10,PB11}.
Note that $\gamma_{\mathrm{r}}^{-1}$ and $\gamma^{-1}$ 
are related to decay rates and that the time period over which
coherence is observed is therefore much longer\cite{PB11}. 
For example, electronic coherences are observed to persist in PC645
at room temperature for times up to 400 fs, a result predicted by
this model as well. Similarly, the time scales over which coherences
persist for FMO, predicted from this model,
are in agreement with all other (far more complex) computations, and with
the earlier experimental results in Ref. \citenum{EC&07}. 

\begin{figure}
\begin{center}
\includegraphics[width = 0.9\columnwidth]{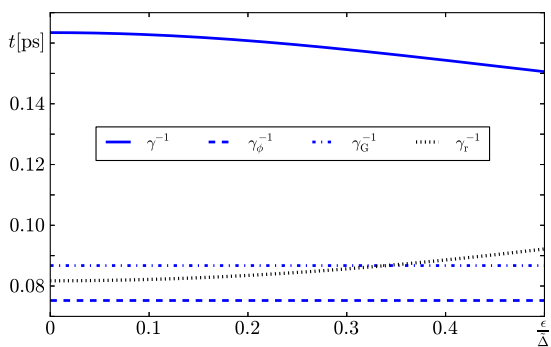}
\caption{
Decoherence time $\gamma^{-1}$ (continuous blue line) as functions
of $\epsilon/2\Delta$, for fixed $\Delta= 35$ cm$^{-1}$ and
at $T=77$~K.
The dashed blue line denotes the dephasing rate used in 
Refs.~\citenum{GM08,RM&09,PH&10} given by $\gamma_{\phi} =
2\pi (k_{\mathrm{B}} T/\hbar)\lambda/\hbar \omega_{\mathrm{c}}$ while the
dotdashed blue line denotes the de\-coherence time
$\tau_{\mathrm{G}} = \sqrt{\hbar^2/2\lambda k_{\mathrm{B}} T}$
used in references \citenum{HR04,GM08,CF09}. 
The relaxation time $\gamma_{\mathrm{r}}^{-1}$ is depicted by the black dotted 
line. Fixed parameters as in Table~\ref{tab:parametersFMO}.
The data for this plot was taken from Ref.~\citenum{PB11}. Note that color
is oniine, but not in the printed text.
} \label{gammaVSepsilon}
\end{center}
\end{figure}

Hence, the long time scales seen to emerge naturally here
from the system parameters.
Why then were far shorter decoherence time scales originally
expected for these systems?
To see this, note that in molecular systems, dynamics  is often
studied between different electronic eigenstates of the system,
separated by greater than $\sim 10^{4}$~cm$^{-1}$, with  no
coupling between them.
In such cases, the dephasing time from Eqs.~(\ref{equ:gammar}) and
Eq.~(\ref{equ:gamma}) would be extremely short.
By contrast, in the case of photosynthetic complexes, energy
transfer occurs between exciton states that are close in energy and,
additionally, are coupled.
This generates a small value for the ratio $\epsilon/2\Delta$ which in
turn is responsible for longer dephasing times (see Fig.~\ref{gammaVSepsilon}).

Additionally, expressions such as $\tau_{\mathrm{G}}$ [Eq.~(\ref{equ:tGrate})],
which are
often used to estimate rates, are only valid at high temperature,
$k_{\mathrm{B}} T \gg \hbar \omega_{\mathrm{c}}$, and at short
times, $t < \omega_{\mathrm{c}}^{-1}$.
Under conditions when the expression for $\tau_\mathrm{G}$ is valid,
the bath modes can be treated classically \cite{TWM01}, as in
Refs.~\citenum{HR04b} and \citenum{FSB08}.
When one is not in the appropriate regime,  the classical evolution of the
bath underestimates quantum coherence effects \cite{TWM01} because
at low temperatures quantum fluctuations overcome thermal
fluctuations\cite{HR04}.
Hence, estimates based on $\tau_\mathrm{G}$ in Eq.~(\ref{equ:tGrate})
are unreliable.
Moreover, as we already mentioned, $\tau_{\mathrm{G}}$ characterizes
a Gaussian decay, while in the experimental results\cite{EC&07,CW&10,PH&10}
the decay is observed to be exponential.

Similarly, the expression $\gamma_{\phi} =
2\pi (k_{\mathrm{B}} T/\hbar)\lambda/\hbar \omega_{\mathrm{c}}$,
another usual expression for estimating the exponential decay
at high temperature, also provides an inadequate representation of the
true physics  and associated dependence on system and bath parameters.
This expression is derived in the limit $\epsilon\rightarrow 0$, $K\rightarrow 0$
and $\omega_{\mathrm{c}} \rightarrow \infty$, so it is expected to result
in decoherence times that are underestimated and misleading\cite{PB11}.

In Fig.~\ref{gammaVSepsilon}, we have shown the explicit dependence
of the various decay rates in Eqs.~(\ref{equ:Omega})-(\ref{equ:gamma})
for fixed values of the ratio $\epsilon/2\Delta$ using the FMO parameters
at 77~K. We observe that the decay rate increases for large values of
this ratio.
Based on this result and our previous discussion, one can identify three
physical features found to be responsible for long  coherence lifetimes:
(i) \emph{the small energy gap between excitonic states}
(ii) \emph{the small ratio of the energy gap to the coupling
between excitonic states}, and
(iii) \emph{the fact that the molecular characteristics place
the system in an effective low temperature regime, even at ambient
conditions.}

In this framework, the observed long lifetimes arise naturally
and are  not surprisingly long.
This physical picture described above, was subsequently verified
by Cao \emph{et al.}\cite{MW&11} using the Hamiltonian described
in Sec.~\ref{sssSec:EffHam}.
A similar physical picture for the survival of coherences has emerged
independently in other physical systems such as spin charge qubits in
quantum dots\cite{PJ&05,HK&07b} or in electronic excitations in semiconducting
carbon nanotubes \cite{HK&07}.
This provides some robustness to the findings in Ref.~\citenum{PB11}.

\subsection{Coherences and energy transfer}
\label{ssec:DoTheyCont}
The possibility that the coherences observed in the 2D electronic
spectroscopy experiments\cite{LCF07,EC&07,ME&09,CW&10,PH&10}
could be assisting the high efficiency of the photosynthetic process has
been one of the more exciting conjectures in the field.
However, there is now sufficient theoretical evidence from different
groups, such as Whaley \textit{et al.}\cite{HSW10}, Olbrich \textit{et al.}\cite{OK10,SR&11},
Cao \textit{et al.}\cite{WL&10,WL&11,MW&11}, Coker \textit{et al.}\cite{HC11}
and Aspuru-Guzik \textit{et al.}\cite{SR&11},
suggest that the presence
of these coherences has only a slight effect on energy
transfer.

\section{Some Future Directions}

Results discussed above apply to the specific  case of EET in photosynthetic
light harvesting. We remark on a number of possible related extensions of
interest.

1. It is worth noting a critical and interesting overview on EET
modelling in light-harvesting complexes
by Huo and Coker in Ref.~\citenum{HC11}.
They point out that despite the complexity, diversity and variability of
the protein-pigment complexes (see Sec.~\ref{sSec:MopDes}), all of
these complexes are highly efficient and the Hamiltonians of these systems, parametrized
using the molecular modelling methodologies discussed in
Sec.~\ref{Sec:MolMod}, share some generic features\cite{HC11}:
(i) \emph{Clusters of chromophores with closely spaced excitation energies
that have appreciable electronic couplings between the cluster members.}
(ii)
\emph{Chromophore states whose excitation energies are isolated,
but which exhibit appreciable electronic coupling to neighboring states.}
(iii)
\emph{These isolated but coupled states are often arranged energetically in
cascade or barrier patterns that channel the directional flow of
energy through these multichromophore networks, and toward reaction
centers.}

These features, and those discussed in Sec.~\ref{sSec:WDoTLiveSoL},
could be used as the target properties for artificially designed light-harvesting
complexes.
These characteristics could be the desirable outcome of molecular
designs based on current or new molecular modelling strategies,
such as those described in Sec.~\ref{Sec:MolMod}.
Work in this direction is worth considering, mainly
because it would relate the studies  above to the ultimate goal of designing
artificial light-harvesting devices inspired by nature, and
currently of great interest (e.g., Ref.  \citenum{BT&11}) .

2. The physical insights discussed in Sec.~\ref{sSec:WDoTLiveSoL}
describe how coherence can be preserved after a single
photon excitation.
Recently, motivated by possible natural scenarios such as
photo-protection\cite{vVv00}, there is interest in the possibility
that the observed coherence can be restored continuously in time\cite{HIW11}.
This situation suggests a role for non-equilibrium phenomena in EET.

The importance of non-equilibrium contributions
can be examined in terms of the results derived, in a different
context, by Pach\'on \textit{et al.} in Ref.~\citenum{GPZ10}.
They have shown that for out-of-equilibrium quantum systems, the border
between the classical and the quantum realms is more intricate than
that for the equilibrium situation.
In particular, they have shown that some quantum features, such as
entanglement, can survive under higher temperatures in non-equilibrium cases
than in the equilibrium case\cite{GPZ10} (see
also Ref.~\citenum{Ved10}).
However, even if the process described in Ref.~\citenum{GPZ10} could
protect the coherences, it is not clear whether or not they
could optimize EET, even though preliminary results show that these
coherence could be spatially directing energy transfer\cite{HIW11}.

3. The results described above show that the notion of a ``high temperature
regime" is relative, it depends on the energetics of the complex and as well
as on the solvent properties.
That is, room temperature could be low temperature for some systems but
high temperature for others.
The fact that EET in photosynthetic complexes relies on the moderate/low
temperature can be seen to be a way of protecting the possible coherent dynamics
that could take place in the light harvesting process.
The reason for this is that, at low temperatures and in the long
time regime, the decay of the correlations is expected to be algebraic,
$1/t^2$, rather than exponential, $\exp(-\gamma_m t)$,\cite{GWT84,HR85,GSI88}.
Thus, in the low temperature regime the coherences are expected to decay
more slowly than in the high temperature regime.
Our discussion in Sec.~\ref{sSec:WDoTLiveSoL} offers the basis for exploring
the design or the search of natural complexes in an effective low temperature,
where the electronic coherences could live even longer than seen at
present.

4. Although methodologies such as  QUAPI \cite{TE&09,NI&11,NBT11}
or the second order cumulant expansion  \cite{IF09,IF09a,IC&10} have provided 
valuable insights into EET, we call attention to some other promising new methodologies
 based
on their reliability over a wide spectrum of parameters, their performance,
and efficiency. They are also
versatile, applying to
different kinds of coupling terms and spectral densities. Specifically, we
are referring to methodologies such PLDM (cf. Sec.~\ref{ssSec:LinApp}) and
those\cite{RR&11,RR&11b,MZC12} based on quantum state diffusion\cite{YDG&99,SG02}
that are worth considering beyond the context
of EET in photosynthetic complexes, e.g., in superconducting carbon nanotubes\cite{HK&07}
or other physical solid-state physics systems\cite{JRS65,GJ&72,SKJ02,MD&95}.
After completing this Perspective, we noted the current interest
in applying and developing efficient methodologies based on density-matrix renormalization-group 
approaches (cf. Ref~\citenum{MRS02,PC&10}), which seem to be highly efficient and could
be used to explore possible mechanisms assisting EET in regions of the parameter 
space beyond the scope of our analysis (cf. Ref.~\citenum{CP&12}) or with highly structured
spectral densities.

\section*{Acknowledgments}
We acknowledge helpful comments on this manuscript by Dr. Torsten Scholak,
and by Professors David F. Coker and Gregory D. Scholes.
This work was supported by the US Air Force Office of Scientific Research
under contract number FA9550-10-1-0260 and by \emph{Comit\'e para el
Desarrollo de la Investigaci\'on} --CODI-- of Universidad de Antioquia, Colombia.

\footnotesize{
\bibliography{pccpMPSv2} 

\providecommand*{\mcitethebibliography}{\thebibliography}
\csname @ifundefined\endcsname{endmcitethebibliography}
{\let\endmcitethebibliography\endthebibliography}{}
\begin{mcitethebibliography}{150}
\providecommand*{\natexlab}[1]{#1}
\providecommand*{\mciteSetBstSublistMode}[1]{}
\providecommand*{\mciteSetBstMaxWidthForm}[2]{}
\providecommand*{\mciteBstWouldAddEndPuncttrue}
  {\def\EndOfBibitem{\unskip.}}
\providecommand*{\mciteBstWouldAddEndPunctfalse}
  {\let\EndOfBibitem\relax}
\providecommand*{\mciteSetBstMidEndSepPunct}[3]{}
\providecommand*{\mciteSetBstSublistLabelBeginEnd}[3]{}
\providecommand*{\EndOfBibitem}{}
\mciteSetBstSublistMode{f}
\mciteSetBstMaxWidthForm{subitem}
{(\emph{\alph{mcitesubitemcount}})}
\mciteSetBstSublistLabelBeginEnd{\mcitemaxwidthsubitemform\space}
{\relax}{\relax}

\bibitem[Jortner \emph{et~al.}(1965)Jortner, Rice, and Silbey]{JRS65}
J.~Jortner, S.~A. Rice and R.~Silbey, in \emph{Excitons and Energy Transfer in
  Molecular Crystals}, ed. O.~Sinanoglu, Academic Press, New York, 1965, p.
  139\relax
\mciteBstWouldAddEndPuncttrue
\mciteSetBstMidEndSepPunct{\mcitedefaultmidpunct}
{\mcitedefaultendpunct}{\mcitedefaultseppunct}\relax
\EndOfBibitem
\bibitem[Gedanken \emph{et~al.}(1972)Gedanken, Jortner, Raz, and
  Sz\"{o}ke]{GJ&72}
A.~Gedanken, J.~Jortner, B.~Raz and A.~Sz\"{o}ke, \emph{J. Chem. Phys.}, 1972,
  \textbf{57}, 3456--3469\relax
\mciteBstWouldAddEndPuncttrue
\mciteSetBstMidEndSepPunct{\mcitedefaultmidpunct}
{\mcitedefaultendpunct}{\mcitedefaultseppunct}\relax
\EndOfBibitem
\bibitem[Schwentner \emph{et~al.}(1985)Schwentner, Koch, and Jortner]{SKJ02}
N.~Schwentner, E.-E. Koch and J.~Jortner, \emph{Electronic Excitations in
  Condensed Rare Gases}, Springer Verlag, Berlin-Heidelberg-New York, 1985,
  vol. 107\relax
\mciteBstWouldAddEndPuncttrue
\mciteSetBstMidEndSepPunct{\mcitedefaultmidpunct}
{\mcitedefaultendpunct}{\mcitedefaultseppunct}\relax
\EndOfBibitem
\bibitem[Moll \emph{et~al.}(1995)Moll, Daehne, Durrant, and Wiersma]{MD&95}
J.~Moll, S.~Daehne, J.~R. Durrant and D.~A. Wiersma, \emph{J. Chem. Phys.},
  1995, \textbf{102}, 6362--6370\relax
\mciteBstWouldAddEndPuncttrue
\mciteSetBstMidEndSepPunct{\mcitedefaultmidpunct}
{\mcitedefaultendpunct}{\mcitedefaultseppunct}\relax
\EndOfBibitem
\bibitem[Renger \emph{et~al.}(2001)Renger, May, and K\"uhn]{RMK01}
T.~Renger, V.~May and O.~K\"uhn, \emph{Phys. Rep.}, 2001, \textbf{343}, 137 --
  254\relax
\mciteBstWouldAddEndPuncttrue
\mciteSetBstMidEndSepPunct{\mcitedefaultmidpunct}
{\mcitedefaultendpunct}{\mcitedefaultseppunct}\relax
\EndOfBibitem
\bibitem[May and K\"uhn(2011)]{MK11}
V.~May and O.~K\"uhn, \emph{Charge and energy transfer dynamics in molecular
  systems}, Wiley-VCH, Weinheim, 3rd edn, 2011\relax
\mciteBstWouldAddEndPuncttrue
\mciteSetBstMidEndSepPunct{\mcitedefaultmidpunct}
{\mcitedefaultendpunct}{\mcitedefaultseppunct}\relax
\EndOfBibitem
\bibitem[Sundstr\"om \emph{et~al.}(1999)Sundstr\"om, Pullerits, and van
  Grondelle]{SPv99}
V.~Sundstr\"om, T.~Pullerits and R.~van Grondelle, \emph{J. Phys. Chem. B},
  1999, \textbf{103}, 2327--2346\relax
\mciteBstWouldAddEndPuncttrue
\mciteSetBstMidEndSepPunct{\mcitedefaultmidpunct}
{\mcitedefaultendpunct}{\mcitedefaultseppunct}\relax
\EndOfBibitem
\bibitem[van Amerongen \emph{et~al.}(2000)van Amerongen, Valkunas, and van
  Grondelle]{vVv00}
H.~van Amerongen, L.~Valkunas and R.~van Grondelle, \emph{Photosynthetic
  Excitons}, World Scientific, Singapore, 2000\relax
\mciteBstWouldAddEndPuncttrue
\mciteSetBstMidEndSepPunct{\mcitedefaultmidpunct}
{\mcitedefaultendpunct}{\mcitedefaultseppunct}\relax
\EndOfBibitem
\bibitem[Salverda \emph{et~al.}(2003)Salverda, Vengris, Krueger, Scholes,
  Czarnoleski, Novoderezhkin, van Amerongen, and van Grondelle]{SV&03}
J.~M. Salverda, M.~Vengris, B.~P. Krueger, G.~D. Scholes, A.~R. Czarnoleski,
  V.~Novoderezhkin, H.~van Amerongen and R.~van Grondelle, \emph{Biophys. J.},
  2003, \textbf{84}, 450--465\relax
\mciteBstWouldAddEndPuncttrue
\mciteSetBstMidEndSepPunct{\mcitedefaultmidpunct}
{\mcitedefaultendpunct}{\mcitedefaultseppunct}\relax
\EndOfBibitem
\bibitem[Lee \emph{et~al.}(2007)Lee, Cheng, and Fleming]{LCF07}
H.~Lee, Y.-C. Cheng and G.~R. Fleming, \emph{Science}, 2007, \textbf{316},
  1462--1465\relax
\mciteBstWouldAddEndPuncttrue
\mciteSetBstMidEndSepPunct{\mcitedefaultmidpunct}
{\mcitedefaultendpunct}{\mcitedefaultseppunct}\relax
\EndOfBibitem
\bibitem[Engel \emph{et~al.}(2007)Engel, Calhoun, Read, Ahn, Man\v{c}al, Cheng,
  Blankenship, and Fleming]{EC&07}
G.~S. Engel, T.~R. Calhoun, E.~L. Read, T.-K. Ahn, T.~Man\v{c}al, Y.-C. Cheng,
  R.~E. Blankenship and G.~R. Fleming, \emph{Nature}, 2007, \textbf{446},
  782--786\relax
\mciteBstWouldAddEndPuncttrue
\mciteSetBstMidEndSepPunct{\mcitedefaultmidpunct}
{\mcitedefaultendpunct}{\mcitedefaultseppunct}\relax
\EndOfBibitem
\bibitem[Mercer \emph{et~al.}(2009)Mercer, El-Taha, Kajumba, Marangos, Tisch,
  Gabrielsen, Cogdell, Springate, and Turcu]{ME&09}
I.~P. Mercer, Y.~C. El-Taha, N.~Kajumba, J.~P. Marangos, J.~W.~G. Tisch,
  M.~Gabrielsen, R.~J. Cogdell, E.~Springate and E.~Turcu, \emph{Phys. Rev.
  Lett.}, 2009, \textbf{102}, 057402\relax
\mciteBstWouldAddEndPuncttrue
\mciteSetBstMidEndSepPunct{\mcitedefaultmidpunct}
{\mcitedefaultendpunct}{\mcitedefaultseppunct}\relax
\EndOfBibitem
\bibitem[Collini \emph{et~al.}(2010)Collini, Wong, Wilk, Curmi, Brumer, and
  Scholes]{CW&10}
E.~Collini, C.~Y. Wong, K.~E. Wilk, P.~M.~G. Curmi, P.~Brumer and G.~D.
  Scholes, \emph{Nature}, 2010, \textbf{463}, 644--647\relax
\mciteBstWouldAddEndPuncttrue
\mciteSetBstMidEndSepPunct{\mcitedefaultmidpunct}
{\mcitedefaultendpunct}{\mcitedefaultseppunct}\relax
\EndOfBibitem
\bibitem[Panitchayangkoon \emph{et~al.}(2010)Panitchayangkoon, Hayes, Fransted,
  Carama, Harel, Wen, Blankenship, and Engel]{PH&10}
G.~Panitchayangkoon, D.~Hayes, K.~A. Fransted, J.~R. Carama, E.~Harel, J.~Wen,
  R.~E. Blankenship and G.~S. Engel, \emph{Proc. Natl. Acad. Sci. USA}, 2010,
  \textbf{107}, 12766--12770\relax
\mciteBstWouldAddEndPuncttrue
\mciteSetBstMidEndSepPunct{\mcitedefaultmidpunct}
{\mcitedefaultendpunct}{\mcitedefaultseppunct}\relax
\EndOfBibitem
\bibitem[Cheng and Fleming(2009)]{CF09}
Y.-C. Cheng and G.~R. Fleming, \emph{Ann. Rev. Phys. Chem.}, 2009, \textbf{60},
  241--262\relax
\mciteBstWouldAddEndPuncttrue
\mciteSetBstMidEndSepPunct{\mcitedefaultmidpunct}
{\mcitedefaultendpunct}{\mcitedefaultseppunct}\relax
\EndOfBibitem
\bibitem[Pach\'on and Brumer(2011)]{PB11}
L.~A. Pach\'on and P.~Brumer, \emph{J. Phys. Chem. Lett.}, 2011, \textbf{2},
  2728--2732\relax
\mciteBstWouldAddEndPuncttrue
\mciteSetBstMidEndSepPunct{\mcitedefaultmidpunct}
{\mcitedefaultendpunct}{\mcitedefaultseppunct}\relax
\EndOfBibitem
\bibitem[Novoderezhkin and van Grondelle(2010)]{Nv10}
V.~I. Novoderezhkin and R.~van Grondelle, \emph{Phys. Chem. Chem. Phys.}, 2010,
  \textbf{12}, 7352--7365\relax
\mciteBstWouldAddEndPuncttrue
\mciteSetBstMidEndSepPunct{\mcitedefaultmidpunct}
{\mcitedefaultendpunct}{\mcitedefaultseppunct}\relax
\EndOfBibitem
\bibitem[Ishizaki \emph{et~al.}(2010)Ishizaki, Calhoun, Schlau-Cohen, and
  Fleming]{IC&10}
A.~Ishizaki, T.~R. Calhoun, G.~S. Schlau-Cohen and G.~R. Fleming, \emph{Phys.
  Chem. Chem. Phys.}, 2010, \textbf{12}, 7319--7337\relax
\mciteBstWouldAddEndPuncttrue
\mciteSetBstMidEndSepPunct{\mcitedefaultmidpunct}
{\mcitedefaultendpunct}{\mcitedefaultseppunct}\relax
\EndOfBibitem
\bibitem[Olaya-Castro and Scholes(2011)]{OS11}
A.~Olaya-Castro and G.~D. Scholes, \emph{Internatl. Rev. Phys. Chem.}, 2011,
  \textbf{30}, 49--77\relax
\mciteBstWouldAddEndPuncttrue
\mciteSetBstMidEndSepPunct{\mcitedefaultmidpunct}
{\mcitedefaultendpunct}{\mcitedefaultseppunct}\relax
\EndOfBibitem
\bibitem[Mennucci and Curutchet(2011)]{MC11}
B.~Mennucci and C.~Curutchet, \emph{Phys. Chem. Chem. Phys.}, 2011,
  \textbf{13}, 11538--11550\relax
\mciteBstWouldAddEndPuncttrue
\mciteSetBstMidEndSepPunct{\mcitedefaultmidpunct}
{\mcitedefaultendpunct}{\mcitedefaultseppunct}\relax
\EndOfBibitem
\bibitem[Schlau-Cohen \emph{et~al.}(2011)Schlau-Cohen, Ishizaki, and
  Fleming]{SIF11}
G.~S. Schlau-Cohen, A.~Ishizaki and G.~R. Fleming, \emph{Chem. Phys.}, 2011,
  \textbf{386}, 1 -- 22\relax
\mciteBstWouldAddEndPuncttrue
\mciteSetBstMidEndSepPunct{\mcitedefaultmidpunct}
{\mcitedefaultendpunct}{\mcitedefaultseppunct}\relax
\EndOfBibitem
\bibitem[K\"onig and Neugebauer(2012)]{KN11}
C.~K\"onig and J.~Neugebauer, \emph{ChemPhysChem.}, 2012, \textbf{13},
  386--425\relax
\mciteBstWouldAddEndPuncttrue
\mciteSetBstMidEndSepPunct{\mcitedefaultmidpunct}
{\mcitedefaultendpunct}{\mcitedefaultseppunct}\relax
\EndOfBibitem
\bibitem[{Th.~F\"orster}(1948)]{Fos48}
{Th.~F\"orster}, \emph{Annalen der Physik}, 1948, \textbf{437}, 55--75\relax
\mciteBstWouldAddEndPuncttrue
\mciteSetBstMidEndSepPunct{\mcitedefaultmidpunct}
{\mcitedefaultendpunct}{\mcitedefaultseppunct}\relax
\EndOfBibitem
\bibitem[{Th.~F\"orster}(1965)]{Fos65}
{Th.~F\"orster}, in \emph{Delocalized excitation and excitation transfer}, ed.
  O.~Sinanoglu, Academic Press, New York, 1965, p.~93\relax
\mciteBstWouldAddEndPuncttrue
\mciteSetBstMidEndSepPunct{\mcitedefaultmidpunct}
{\mcitedefaultendpunct}{\mcitedefaultseppunct}\relax
\EndOfBibitem
\bibitem[Redfield(1957)]{Red57}
A.~G. Redfield, \emph{IBM J. Res. Dev.}, 1957, \textbf{1}, 19--31\relax
\mciteBstWouldAddEndPuncttrue
\mciteSetBstMidEndSepPunct{\mcitedefaultmidpunct}
{\mcitedefaultendpunct}{\mcitedefaultseppunct}\relax
\EndOfBibitem
\bibitem[Olaya-Castro \emph{et~al.}(2008)Olaya-Castro, Lee, Olsen, and
  Johnson]{OL&08}
A.~Olaya-Castro, C.~F. Lee, F.~F. Olsen and N.~F. Johnson, \emph{Phys. Rev. B},
  2008, \textbf{78}, 085115\relax
\mciteBstWouldAddEndPuncttrue
\mciteSetBstMidEndSepPunct{\mcitedefaultmidpunct}
{\mcitedefaultendpunct}{\mcitedefaultseppunct}\relax
\EndOfBibitem
\bibitem[Plenio and Huelga(2008)]{PH08}
M.~B. Plenio and S.~F. Huelga, \emph{New J. Phys.}, 2008, \textbf{10},
  113019\relax
\mciteBstWouldAddEndPuncttrue
\mciteSetBstMidEndSepPunct{\mcitedefaultmidpunct}
{\mcitedefaultendpunct}{\mcitedefaultseppunct}\relax
\EndOfBibitem
\bibitem[Rebentrost \emph{et~al.}(2009)Rebentrost, Mohseni, Kassal, Lloyd, and
  Aspuru-Guzik]{RM&09}
P.~Rebentrost, M.~Mohseni, I.~Kassal, S.~Lloyd and A.~Aspuru-Guzik, \emph{New
  J. Phys.}, 2009, \textbf{11}, 033003\relax
\mciteBstWouldAddEndPuncttrue
\mciteSetBstMidEndSepPunct{\mcitedefaultmidpunct}
{\mcitedefaultendpunct}{\mcitedefaultseppunct}\relax
\EndOfBibitem
\bibitem[Huo and Coker(2010)]{HC10}
P.~Huo and D.~F. Coker, \emph{J. Chem. Phys.}, 2010, \textbf{133}, 184108\relax
\mciteBstWouldAddEndPuncttrue
\mciteSetBstMidEndSepPunct{\mcitedefaultmidpunct}
{\mcitedefaultendpunct}{\mcitedefaultseppunct}\relax
\EndOfBibitem
\bibitem[Tao and Miller(2010)]{TM10}
G.~Tao and W.~H. Miller, \emph{J. Phys. Chem. Lett.}, 2010, \textbf{1},
  891--894\relax
\mciteBstWouldAddEndPuncttrue
\mciteSetBstMidEndSepPunct{\mcitedefaultmidpunct}
{\mcitedefaultendpunct}{\mcitedefaultseppunct}\relax
\EndOfBibitem
\bibitem[Huo and Coker(2011)]{HC11}
P.~Huo and D.~F. Coker, \emph{J. Phys. Chem. Lett.}, 2011, \textbf{2},
  825--833\relax
\mciteBstWouldAddEndPuncttrue
\mciteSetBstMidEndSepPunct{\mcitedefaultmidpunct}
{\mcitedefaultendpunct}{\mcitedefaultseppunct}\relax
\EndOfBibitem
\bibitem[Huo and Coker(2011)]{HC11b}
P.~Huo and D.~F. Coker, \emph{J. Chem. Phys.}, 2011, \textbf{135}, 201101\relax
\mciteBstWouldAddEndPuncttrue
\mciteSetBstMidEndSepPunct{\mcitedefaultmidpunct}
{\mcitedefaultendpunct}{\mcitedefaultseppunct}\relax
\EndOfBibitem
\bibitem[Moix \emph{et~al.}(2011)Moix, Wu, Huo, Coker, and Cao]{MW&11}
J.~Moix, J.~Wu, P.~Huo, D.~Coker and J.~Cao, \emph{J. Phys. Chem. Lett.}, 2011,
  \textbf{2}, 3045--3052\relax
\mciteBstWouldAddEndPuncttrue
\mciteSetBstMidEndSepPunct{\mcitedefaultmidpunct}
{\mcitedefaultendpunct}{\mcitedefaultseppunct}\relax
\EndOfBibitem
\bibitem[Nalbach \emph{et~al.}(2011)Nalbach, Braun, and Thorwart]{NBT11}
P.~Nalbach, D.~Braun and M.~Thorwart, \emph{Phys. Rev. E}, 2011, \textbf{84},
  041926\relax
\mciteBstWouldAddEndPuncttrue
\mciteSetBstMidEndSepPunct{\mcitedefaultmidpunct}
{\mcitedefaultendpunct}{\mcitedefaultseppunct}\relax
\EndOfBibitem
\bibitem[Shim \emph{et~al.}()Shim, Rebentrost, Valleau, and
  Aspuru-Guzik]{SR&11}
S.~Shim, P.~Rebentrost, S.~Valleau and A.~Aspuru-Guzik, \emph{Boiphys. J.},
  \textbf{102}, 649--660\relax
\mciteBstWouldAddEndPuncttrue
\mciteSetBstMidEndSepPunct{\mcitedefaultmidpunct}
{\mcitedefaultendpunct}{\mcitedefaultseppunct}\relax
\EndOfBibitem
\bibitem[Scholes(2010)]{Sch10}
G.~D. Scholes, \emph{The Journal of Physical Chemistry Letters}, 2010,
  \textbf{1}, 2--8\relax
\mciteBstWouldAddEndPuncttrue
\mciteSetBstMidEndSepPunct{\mcitedefaultmidpunct}
{\mcitedefaultendpunct}{\mcitedefaultseppunct}\relax
\EndOfBibitem
\bibitem[Adolphs and Renger(2006)]{AR06}
J.~Adolphs and T.~Renger, \emph{Biophys. J.}, 2006, \textbf{91},
  2778--2797\relax
\mciteBstWouldAddEndPuncttrue
\mciteSetBstMidEndSepPunct{\mcitedefaultmidpunct}
{\mcitedefaultendpunct}{\mcitedefaultseppunct}\relax
\EndOfBibitem
\bibitem[Tronrud \emph{et~al.}(2009)Tronrud, Wen, Gay, and Blankenship]{TW&09}
D.~Tronrud, J.~Wen, L.~Gay and R.~Blankenship, \emph{Photosynth. Res.}, 2009,
  \textbf{100}, 79--87\relax
\mciteBstWouldAddEndPuncttrue
\mciteSetBstMidEndSepPunct{\mcitedefaultmidpunct}
{\mcitedefaultendpunct}{\mcitedefaultseppunct}\relax
\EndOfBibitem
\bibitem[Schmidt~am Busch \emph{et~al.}(2011)Schmidt~am Busch, M\"uh,
  El-Amine~Madjet, and Renger]{SM&11}
M.~Schmidt~am Busch, F.~M\"uh, M.~El-Amine~Madjet and T.~Renger, \emph{J. Phys.
  Chem. Lett.}, 2011, \textbf{2}, 93--98\relax
\mciteBstWouldAddEndPuncttrue
\mciteSetBstMidEndSepPunct{\mcitedefaultmidpunct}
{\mcitedefaultendpunct}{\mcitedefaultseppunct}\relax
\EndOfBibitem
\bibitem[Mirkovic \emph{et~al.}(2007)Mirkovic, Doust, Kim, Wilk, Curutchet,
  Mennucci, Cammi, Curmi, and Scholes]{MD&07}
T.~Mirkovic, A.~B. Doust, J.~Kim, K.~E. Wilk, C.~Curutchet, B.~Mennucci,
  R.~Cammi, P.~M.~G. Curmi and G.~D. Scholes, \emph{Photochem. Photobiol.
  Sci.}, 2007, \textbf{6"}, 964--975\relax
\mciteBstWouldAddEndPuncttrue
\mciteSetBstMidEndSepPunct{\mcitedefaultmidpunct}
{\mcitedefaultendpunct}{\mcitedefaultseppunct}\relax
\EndOfBibitem
\bibitem[Ratner(1990)]{Rat90}
M.~A. Ratner, \emph{J. Phys. Chem.}, 1990, \textbf{94}, 4877--4883\relax
\mciteBstWouldAddEndPuncttrue
\mciteSetBstMidEndSepPunct{\mcitedefaultmidpunct}
{\mcitedefaultendpunct}{\mcitedefaultseppunct}\relax
\EndOfBibitem
\bibitem[Prezhdo and Rossky(1997)]{PR97}
O.~V. Prezhdo and P.~J. Rossky, \emph{J. Chem. Phys.}, 1997, \textbf{107},
  5863--5878\relax
\mciteBstWouldAddEndPuncttrue
\mciteSetBstMidEndSepPunct{\mcitedefaultmidpunct}
{\mcitedefaultendpunct}{\mcitedefaultseppunct}\relax
\EndOfBibitem
\bibitem[Prezhdo and Rossky(1998)]{PR98}
O.~V. Prezhdo and P.~J. Rossky, \emph{Phys. Rev. Lett.}, 1998, \textbf{81},
  5294--5297\relax
\mciteBstWouldAddEndPuncttrue
\mciteSetBstMidEndSepPunct{\mcitedefaultmidpunct}
{\mcitedefaultendpunct}{\mcitedefaultseppunct}\relax
\EndOfBibitem
\bibitem[Shapiro and Brumer(2011)]{SB11}
M.~Shapiro and P.~Brumer, \emph{Quantum Control of Molecular Processes},
  Wiley-VCH, Weinheim, 2nd edn, 2011\relax
\mciteBstWouldAddEndPuncttrue
\mciteSetBstMidEndSepPunct{\mcitedefaultmidpunct}
{\mcitedefaultendpunct}{\mcitedefaultseppunct}\relax
\EndOfBibitem
\bibitem[Hwang and Rossky(2004)]{HR04b}
H.~Hwang and P.~J. Rossky, \emph{J. Phys. Chem. B}, 2004, \textbf{108},
  6723--6732\relax
\mciteBstWouldAddEndPuncttrue
\mciteSetBstMidEndSepPunct{\mcitedefaultmidpunct}
{\mcitedefaultendpunct}{\mcitedefaultseppunct}\relax
\EndOfBibitem
\bibitem[Franco and Brumer(2008)]{FSB08}
I.~Franco and P.~Brumer, \emph{J. Chem. Phys.}, 2008, \textbf{128},
  244905\relax
\mciteBstWouldAddEndPuncttrue
\mciteSetBstMidEndSepPunct{\mcitedefaultmidpunct}
{\mcitedefaultendpunct}{\mcitedefaultseppunct}\relax
\EndOfBibitem
\bibitem[Franco and Brumer(2011)]{FB11}
I.~Franco and P.~Brumer, 2011,  arXiv:1111.6461v1\relax
\mciteBstWouldAddEndPuncttrue
\mciteSetBstMidEndSepPunct{\mcitedefaultmidpunct}
{\mcitedefaultendpunct}{\mcitedefaultseppunct}\relax
\EndOfBibitem
\bibitem[Schlosshauer(2007)]{Sch07}
M.~Schlosshauer, \emph{Decoherence and the Quantum-To-Classical Transition},
  Springer-Verlag, Berlin, 2007\relax
\mciteBstWouldAddEndPuncttrue
\mciteSetBstMidEndSepPunct{\mcitedefaultmidpunct}
{\mcitedefaultendpunct}{\mcitedefaultseppunct}\relax
\EndOfBibitem
\bibitem[Nazir(2009)]{Naz09}
A.~Nazir, \emph{Phys. Rev. Lett.}, 2009, \textbf{103}, 146404\relax
\mciteBstWouldAddEndPuncttrue
\mciteSetBstMidEndSepPunct{\mcitedefaultmidpunct}
{\mcitedefaultendpunct}{\mcitedefaultseppunct}\relax
\EndOfBibitem
\bibitem[Fassioli \emph{et~al.}(2010)Fassioli, Nazir, and Olaya-Castro]{FNO10}
F.~Fassioli, A.~Nazir and A.~Olaya-Castro, \emph{J. Phys. Chem. Lett.}, 2010,
  \textbf{1}, 2139--2143\relax
\mciteBstWouldAddEndPuncttrue
\mciteSetBstMidEndSepPunct{\mcitedefaultmidpunct}
{\mcitedefaultendpunct}{\mcitedefaultseppunct}\relax
\EndOfBibitem
\bibitem[Nalbach \emph{et~al.}(2010)Nalbach, Eckel, and Thorwart]{NET10}
P.~Nalbach, J.~Eckel and M.~Thorwart, \emph{New J. Phys.}, 2010, \textbf{12},
  065043\relax
\mciteBstWouldAddEndPuncttrue
\mciteSetBstMidEndSepPunct{\mcitedefaultmidpunct}
{\mcitedefaultendpunct}{\mcitedefaultseppunct}\relax
\EndOfBibitem
\bibitem[Str\"{u}mpfer and Schulten(2011)]{SS11}
J.~Str\"{u}mpfer and K.~Schulten, \emph{J. Chem. Phys.}, 2011, \textbf{134},
  095102\relax
\mciteBstWouldAddEndPuncttrue
\mciteSetBstMidEndSepPunct{\mcitedefaultmidpunct}
{\mcitedefaultendpunct}{\mcitedefaultseppunct}\relax
\EndOfBibitem
\bibitem[Caldeira and Leggett(1983)]{CL83}
A.~O. Caldeira and A.~L. Leggett, \emph{Physica A}, 1983, \textbf{121},
  587\relax
\mciteBstWouldAddEndPuncttrue
\mciteSetBstMidEndSepPunct{\mcitedefaultmidpunct}
{\mcitedefaultendpunct}{\mcitedefaultseppunct}\relax
\EndOfBibitem
\bibitem[Grabert \emph{et~al.}(1988)Grabert, Schramm, and Ingold]{GSI88}
H.~Grabert, P.~Schramm and G.~L. Ingold, \emph{Phys. Rep.}, 1988, \textbf{168},
  115\relax
\mciteBstWouldAddEndPuncttrue
\mciteSetBstMidEndSepPunct{\mcitedefaultmidpunct}
{\mcitedefaultendpunct}{\mcitedefaultseppunct}\relax
\EndOfBibitem
\bibitem[Ingold(2002)]{Ing02}
G.-L. Ingold, \emph{Coherent Evolution in Noisy Environments}, Springer Verlag,
  Berlin-Heidelberg-New York, 2002, vol. 611\relax
\mciteBstWouldAddEndPuncttrue
\mciteSetBstMidEndSepPunct{\mcitedefaultmidpunct}
{\mcitedefaultendpunct}{\mcitedefaultseppunct}\relax
\EndOfBibitem
\bibitem[Weiss(2008)]{Wei08}
U.~Weiss, \emph{Quantum Dissipative Systems}, World Scientific, Singapore, 3rd
  edn, 2008\relax
\mciteBstWouldAddEndPuncttrue
\mciteSetBstMidEndSepPunct{\mcitedefaultmidpunct}
{\mcitedefaultendpunct}{\mcitedefaultseppunct}\relax
\EndOfBibitem
\bibitem[Mukamel(1999)]{Muk99}
S.~Mukamel, \emph{Principles of Nonlinear Optical Spectroscopy}, Oxford
  University Press, New York, 1999\relax
\mciteBstWouldAddEndPuncttrue
\mciteSetBstMidEndSepPunct{\mcitedefaultmidpunct}
{\mcitedefaultendpunct}{\mcitedefaultseppunct}\relax
\EndOfBibitem
\bibitem[Olbrich \emph{et~al.}(2011)Olbrich, Str\"umpfer, Schulten, and
  Kleinekath\"ofer]{OS&11a}
C.~Olbrich, J.~Str\"umpfer, K.~Schulten and U.~Kleinekath\"ofer, \emph{J. Phys.
  Chem. Lett.}, 2011, \textbf{2}, 1771--1776\relax
\mciteBstWouldAddEndPuncttrue
\mciteSetBstMidEndSepPunct{\mcitedefaultmidpunct}
{\mcitedefaultendpunct}{\mcitedefaultseppunct}\relax
\EndOfBibitem
\bibitem[Gilmore and McKenzie(2005)]{GM05}
J.~B. Gilmore and R.~H. McKenzie, \emph{J. Phys. Condens. Matter}, 2005,
  \textbf{17}, 1735--1746\relax
\mciteBstWouldAddEndPuncttrue
\mciteSetBstMidEndSepPunct{\mcitedefaultmidpunct}
{\mcitedefaultendpunct}{\mcitedefaultseppunct}\relax
\EndOfBibitem
\bibitem[Leggett \emph{et~al.}(1987)Leggett, Chakravarty, Dorsey, Fisher, Garg,
  and Zwerger]{LC&87}
A.~J. Leggett, S.~Chakravarty, A.~T. Dorsey, M.~P.~A. Fisher, A.~Garg and
  W.~Zwerger, \emph{Rev. Mod. Phys.}, 1987, \textbf{59}, 1--85\relax
\mciteBstWouldAddEndPuncttrue
\mciteSetBstMidEndSepPunct{\mcitedefaultmidpunct}
{\mcitedefaultendpunct}{\mcitedefaultseppunct}\relax
\EndOfBibitem
\bibitem[Gilmore and McKenzie(2006)]{GM06}
J.~B. Gilmore and R.~H. McKenzie, \emph{Chem. Phys. Lett.}, 2006, \textbf{421},
  266\relax
\mciteBstWouldAddEndPuncttrue
\mciteSetBstMidEndSepPunct{\mcitedefaultmidpunct}
{\mcitedefaultendpunct}{\mcitedefaultseppunct}\relax
\EndOfBibitem
\bibitem[Nemeth \emph{et~al.}(2008)Nemeth, Milota, Man\v{c}al, Luke\v{s},
  Kauffmann, and Sperling]{NM&08}
A.~Nemeth, F.~Milota, T.~Man\v{c}al, V.~Luke\v{s}, H.~F. Kauffmann and
  J.~Sperling, \emph{Chem. Phys. Lett.}, 2008, \textbf{459}, 94 -- 99\relax
\mciteBstWouldAddEndPuncttrue
\mciteSetBstMidEndSepPunct{\mcitedefaultmidpunct}
{\mcitedefaultendpunct}{\mcitedefaultseppunct}\relax
\EndOfBibitem
\bibitem[Jiang and Brumer(1991)]{JB91}
X.-P. Jiang and P.~Brumer, \emph{J. Chem. Phys.}, 1991, \textbf{94}, 5833\relax
\mciteBstWouldAddEndPuncttrue
\mciteSetBstMidEndSepPunct{\mcitedefaultmidpunct}
{\mcitedefaultendpunct}{\mcitedefaultseppunct}\relax
\EndOfBibitem
\bibitem[Hoki and Brumer(2009)]{HB07}
H.~Hoki and P.~Brumer, \emph{Chem. Phys. Lett.}, 2009, \textbf{468}, 27\relax
\mciteBstWouldAddEndPuncttrue
\mciteSetBstMidEndSepPunct{\mcitedefaultmidpunct}
{\mcitedefaultendpunct}{\mcitedefaultseppunct}\relax
\EndOfBibitem
\bibitem[Man\v{c}al and Valkunas(2010)]{MV10}
T.~Man\v{c}al and L.~Valkunas, \emph{New J. Phys.}, 2010, \textbf{12},
  065044\relax
\mciteBstWouldAddEndPuncttrue
\mciteSetBstMidEndSepPunct{\mcitedefaultmidpunct}
{\mcitedefaultendpunct}{\mcitedefaultseppunct}\relax
\EndOfBibitem
\bibitem[Hoki and Brumer(2011)]{HB11}
H.~Hoki and P.~Brumer, \emph{Procedia Chem.}, 2011, \textbf{3}, 122\relax
\mciteBstWouldAddEndPuncttrue
\mciteSetBstMidEndSepPunct{\mcitedefaultmidpunct}
{\mcitedefaultendpunct}{\mcitedefaultseppunct}\relax
\EndOfBibitem
\bibitem[Brumer and Shapiro(2011)]{BS11}
P.~Brumer and M.~Shapiro, \emph{Molecular Response in One Photon Absorption:
  Coherent Pulsed Laser vs. Thermal Incoherent Source}, 2011,
  arXiv:1109.0026v2\relax
\mciteBstWouldAddEndPuncttrue
\mciteSetBstMidEndSepPunct{\mcitedefaultmidpunct}
{\mcitedefaultendpunct}{\mcitedefaultseppunct}\relax
\EndOfBibitem
\bibitem[Pach\'on and Brumer(2012)]{PB12}
L.~A. Pach\'on and P.~Brumer, \emph{Incoherent Excitation of Open Quantum
  Systems}, 2012,  arXiv:\relax
\mciteBstWouldAddEndPuncttrue
\mciteSetBstMidEndSepPunct{\mcitedefaultmidpunct}
{\mcitedefaultendpunct}{\mcitedefaultseppunct}\relax
\EndOfBibitem
\bibitem[Ford \emph{et~al.}(1985)Ford, Lewis, and O'Connell]{FLO85}
G.~W. Ford, J.~T. Lewis and R.~F. O'Connell, \emph{Phys. Rev. Lett.}, 1985,
  \textbf{55}, 2273\relax
\mciteBstWouldAddEndPuncttrue
\mciteSetBstMidEndSepPunct{\mcitedefaultmidpunct}
{\mcitedefaultendpunct}{\mcitedefaultseppunct}\relax
\EndOfBibitem
\bibitem[Ford \emph{et~al.}(1987)Ford, Lewis, and O'Connell]{FLO87}
G.~W. Ford, J.~T. Lewis and R.~F. O'Connell, \emph{Phys. Rev. A}, 1987,
  \textbf{36}, 1466\relax
\mciteBstWouldAddEndPuncttrue
\mciteSetBstMidEndSepPunct{\mcitedefaultmidpunct}
{\mcitedefaultendpunct}{\mcitedefaultseppunct}\relax
\EndOfBibitem
\bibitem[Ford \emph{et~al.}(1988)Ford, Lewis, and O'Connell]{FLO88}
G.~W. Ford, J.~T. Lewis and R.~F. O'Connell, \emph{Phys. Rev. A}, 1988,
  \textbf{37}, 4419\relax
\mciteBstWouldAddEndPuncttrue
\mciteSetBstMidEndSepPunct{\mcitedefaultmidpunct}
{\mcitedefaultendpunct}{\mcitedefaultseppunct}\relax
\EndOfBibitem
\bibitem[Barone and Caldeira(1991)]{BC91}
P.~M. V.~B. Barone and A.~O. Caldeira, \emph{Phys. Rev. A}, 1991, \textbf{43},
  57\relax
\mciteBstWouldAddEndPuncttrue
\mciteSetBstMidEndSepPunct{\mcitedefaultmidpunct}
{\mcitedefaultendpunct}{\mcitedefaultseppunct}\relax
\EndOfBibitem
\bibitem[Ford and O'Connell(1998)]{FO98}
G.~W. Ford and R.~F. O'Connell, \emph{Phys. Rev. A}, 1998, \textbf{57},
  3112\relax
\mciteBstWouldAddEndPuncttrue
\mciteSetBstMidEndSepPunct{\mcitedefaultmidpunct}
{\mcitedefaultendpunct}{\mcitedefaultseppunct}\relax
\EndOfBibitem
\bibitem[Lindblad(1976)]{Lin76}
G.~Lindblad, \emph{Commun. Math. Phys.}, 1976, \textbf{48}, 119\relax
\mciteBstWouldAddEndPuncttrue
\mciteSetBstMidEndSepPunct{\mcitedefaultmidpunct}
{\mcitedefaultendpunct}{\mcitedefaultseppunct}\relax
\EndOfBibitem
\bibitem[Gorini \emph{et~al.}(1978)Gorini, Frigerio, Verri, Kossakowski, and
  Sudarshan]{GF&78}
V.~Gorini, A.~Frigerio, M.~Verri, A.~Kossakowski and E.~C.~G. Sudarshan,
  \emph{Rep. Math. Phys.}, 1978, \textbf{13}, 149\relax
\mciteBstWouldAddEndPuncttrue
\mciteSetBstMidEndSepPunct{\mcitedefaultmidpunct}
{\mcitedefaultendpunct}{\mcitedefaultseppunct}\relax
\EndOfBibitem
\bibitem[Cheng and Silbey(2005)]{CS05}
Y.~C. Cheng and R.~J. Silbey, \emph{J. Phys. Chem. B}, 2005, \textbf{109},
  21399--21405\relax
\mciteBstWouldAddEndPuncttrue
\mciteSetBstMidEndSepPunct{\mcitedefaultmidpunct}
{\mcitedefaultendpunct}{\mcitedefaultseppunct}\relax
\EndOfBibitem
\bibitem[Ishizaki and Fleming(2009)]{IF09a}
A.~Ishizaki and G.~R. Fleming, \emph{J. Chem. Phys.}, 2009, \textbf{130},
  234111\relax
\mciteBstWouldAddEndPuncttrue
\mciteSetBstMidEndSepPunct{\mcitedefaultmidpunct}
{\mcitedefaultendpunct}{\mcitedefaultseppunct}\relax
\EndOfBibitem
\bibitem[Jang \emph{et~al.}(2004)Jang, Newton, and Silbey]{JNS04}
S.~Jang, M.~D. Newton and R.~J. Silbey, \emph{Phys. Rev. Lett.}, 2004,
  \textbf{92}, 218301\relax
\mciteBstWouldAddEndPuncttrue
\mciteSetBstMidEndSepPunct{\mcitedefaultmidpunct}
{\mcitedefaultendpunct}{\mcitedefaultseppunct}\relax
\EndOfBibitem
\bibitem[Fleming and Cho(1996)]{FC96}
G.~R. Fleming and M.~Cho, \emph{Ann. Rev. Phys. Chem.}, 1996, \textbf{47},
  109--134\relax
\mciteBstWouldAddEndPuncttrue
\mciteSetBstMidEndSepPunct{\mcitedefaultmidpunct}
{\mcitedefaultendpunct}{\mcitedefaultseppunct}\relax
\EndOfBibitem
\bibitem[Breuer \emph{et~al.}(2009)Breuer, Laine, and Piilo]{BLP09}
H.-P. Breuer, E.-M. Laine and J.~Piilo, \emph{Phys. Rev. Lett.}, 2009,
  \textbf{103}, 210401\relax
\mciteBstWouldAddEndPuncttrue
\mciteSetBstMidEndSepPunct{\mcitedefaultmidpunct}
{\mcitedefaultendpunct}{\mcitedefaultseppunct}\relax
\EndOfBibitem
\bibitem[Rivas \emph{et~al.}(2010)Rivas, Huelga, and Plenio]{RHP10}
A.~Rivas, S.~F. Huelga and M.~B. Plenio, \emph{Phys. Rev. Lett.}, 2010,
  \textbf{105}, 050403\relax
\mciteBstWouldAddEndPuncttrue
\mciteSetBstMidEndSepPunct{\mcitedefaultmidpunct}
{\mcitedefaultendpunct}{\mcitedefaultseppunct}\relax
\EndOfBibitem
\bibitem[Pach\'on and Brumer(2012)]{PB12b}
L.~A. Pach\'on and P.~Brumer, 2012,  In preparation\relax
\mciteBstWouldAddEndPuncttrue
\mciteSetBstMidEndSepPunct{\mcitedefaultmidpunct}
{\mcitedefaultendpunct}{\mcitedefaultseppunct}\relax
\EndOfBibitem
\bibitem[Breuer and Petruccione(2002)]{BP02}
H.-P. Breuer and F.~Petruccione, \emph{The Theory of Open Quantum Systems},
  Oxford University Press, Oxford, 2002\relax
\mciteBstWouldAddEndPuncttrue
\mciteSetBstMidEndSepPunct{\mcitedefaultmidpunct}
{\mcitedefaultendpunct}{\mcitedefaultseppunct}\relax
\EndOfBibitem
\bibitem[Ishizaki and Fleming(2009)]{IF09}
A.~Ishizaki and G.~R. Fleming, \emph{Proc. Natl. Acad. Sci. USA}, 2009,
  \textbf{106}, 17255--17260\relax
\mciteBstWouldAddEndPuncttrue
\mciteSetBstMidEndSepPunct{\mcitedefaultmidpunct}
{\mcitedefaultendpunct}{\mcitedefaultseppunct}\relax
\EndOfBibitem
\bibitem[Nakajima(1958)]{Nak58}
S.~Nakajima, \emph{Progr. Theor. Phys.}, 1958, \textbf{20}, 948--959\relax
\mciteBstWouldAddEndPuncttrue
\mciteSetBstMidEndSepPunct{\mcitedefaultmidpunct}
{\mcitedefaultendpunct}{\mcitedefaultseppunct}\relax
\EndOfBibitem
\bibitem[Zwanzig(1960)]{Zwa60}
R.~Zwanzig, \emph{J. Chem. Phys.}, 1960, \textbf{33}, 1338--1341\relax
\mciteBstWouldAddEndPuncttrue
\mciteSetBstMidEndSepPunct{\mcitedefaultmidpunct}
{\mcitedefaultendpunct}{\mcitedefaultseppunct}\relax
\EndOfBibitem
\bibitem[Zwanzig(1961)]{Zwa61}
R.~Zwanzig, \emph{Phys. Rev.}, 1961, \textbf{124}, 983--992\relax
\mciteBstWouldAddEndPuncttrue
\mciteSetBstMidEndSepPunct{\mcitedefaultmidpunct}
{\mcitedefaultendpunct}{\mcitedefaultseppunct}\relax
\EndOfBibitem
\bibitem[Grabert(1982)]{Gra82}
H.~Grabert, \emph{Projection Operator Techniques in Nonequilibrium Statistical
  Mechanics}, Springer Verlag, Berlin-Heidelberg-New York, 1982, vol.~95\relax
\mciteBstWouldAddEndPuncttrue
\mciteSetBstMidEndSepPunct{\mcitedefaultmidpunct}
{\mcitedefaultendpunct}{\mcitedefaultseppunct}\relax
\EndOfBibitem
\bibitem[Singh and Brumer(2011)]{SiB11}
N.~Singh and P.~Brumer, \emph{Faraday Discuss.}, 2011, \textbf{153},
  41--50\relax
\mciteBstWouldAddEndPuncttrue
\mciteSetBstMidEndSepPunct{\mcitedefaultmidpunct}
{\mcitedefaultendpunct}{\mcitedefaultseppunct}\relax
\EndOfBibitem
\bibitem[Shibata and Arimitsu(1980)]{SA80}
F.~Shibata and T.~Arimitsu, \emph{J. Phys. Soc. Jap.}, 1980, \textbf{49},
  891--897\relax
\mciteBstWouldAddEndPuncttrue
\mciteSetBstMidEndSepPunct{\mcitedefaultmidpunct}
{\mcitedefaultendpunct}{\mcitedefaultseppunct}\relax
\EndOfBibitem
\bibitem[Ritschel \emph{et~al.}(2011)Ritschel, Roden, Strunz, and
  Eisfeld]{Str11}
G.~Ritschel, J.~Roden, W.~T. Strunz and A.~Eisfeld, \emph{New J. Phys.}, 2011,
  \textbf{13}, 113034\relax
\mciteBstWouldAddEndPuncttrue
\mciteSetBstMidEndSepPunct{\mcitedefaultmidpunct}
{\mcitedefaultendpunct}{\mcitedefaultseppunct}\relax
\EndOfBibitem
\bibitem[Zhu \emph{et~al.}(2011)Zhu, Kais, Rebentrost, and Aspuru-Guzik]{JS&11}
J.~Zhu, S.~Kais, P.~Rebentrost and A.~Aspuru-Guzik, \emph{J. Phys. Chem. B},
  2011, \textbf{115}, 1531--1537\relax
\mciteBstWouldAddEndPuncttrue
\mciteSetBstMidEndSepPunct{\mcitedefaultmidpunct}
{\mcitedefaultendpunct}{\mcitedefaultseppunct}\relax
\EndOfBibitem
\bibitem[Meyera and Miller(1979)]{MM79}
H.-D. Meyera and W.~H. Miller, \emph{J. Chem. Phys.}, 1979, \textbf{70},
  3214--3223\relax
\mciteBstWouldAddEndPuncttrue
\mciteSetBstMidEndSepPunct{\mcitedefaultmidpunct}
{\mcitedefaultendpunct}{\mcitedefaultseppunct}\relax
\EndOfBibitem
\bibitem[Stock and Thoss(1997)]{ST97}
G.~Stock and M.~Thoss, \emph{Phys. Rev. Lett.}, 1997, \textbf{78},
  578--581\relax
\mciteBstWouldAddEndPuncttrue
\mciteSetBstMidEndSepPunct{\mcitedefaultmidpunct}
{\mcitedefaultendpunct}{\mcitedefaultseppunct}\relax
\EndOfBibitem
\bibitem[Thoss and Stock(1999)]{TS99}
M.~Thoss and G.~Stock, \emph{Phys. Rev. A}, 1999, \textbf{59}, 64--79\relax
\mciteBstWouldAddEndPuncttrue
\mciteSetBstMidEndSepPunct{\mcitedefaultmidpunct}
{\mcitedefaultendpunct}{\mcitedefaultseppunct}\relax
\EndOfBibitem
\bibitem[Bonella \emph{et~al.}(2005)Bonella, Montemayor, and Coker]{BMC05}
S.~Bonella, D.~Montemayor and D.~F. Coker, \emph{Proc. Natl. Acad. Sci. USA},
  2005, \textbf{102}, 6715--6719\relax
\mciteBstWouldAddEndPuncttrue
\mciteSetBstMidEndSepPunct{\mcitedefaultmidpunct}
{\mcitedefaultendpunct}{\mcitedefaultseppunct}\relax
\EndOfBibitem
\bibitem[Bonella and Coker(2005)]{BC05}
S.~Bonella and D.~F. Coker, \emph{J. Chem. Phys.}, 2005, \textbf{122},
  194102\relax
\mciteBstWouldAddEndPuncttrue
\mciteSetBstMidEndSepPunct{\mcitedefaultmidpunct}
{\mcitedefaultendpunct}{\mcitedefaultseppunct}\relax
\EndOfBibitem
\bibitem[Dunkel \emph{et~al.}(2008)Dunkel, Bonella, and Coker]{DBC08}
E.~R. Dunkel, S.~Bonella and D.~F. Coker, \emph{J. Chem. Phys.}, 2008,
  \textbf{129}, 114106\relax
\mciteBstWouldAddEndPuncttrue
\mciteSetBstMidEndSepPunct{\mcitedefaultmidpunct}
{\mcitedefaultendpunct}{\mcitedefaultseppunct}\relax
\EndOfBibitem
\bibitem[Dittrich and Pach\'on(2009)]{DP09}
T.~Dittrich and L.~A. Pach\'on, \emph{Phys. Rev. Lett.}, 2009, \textbf{102},
  150401\relax
\mciteBstWouldAddEndPuncttrue
\mciteSetBstMidEndSepPunct{\mcitedefaultmidpunct}
{\mcitedefaultendpunct}{\mcitedefaultseppunct}\relax
\EndOfBibitem
\bibitem[Dittrich \emph{et~al.}(2010)Dittrich, G\'{o}mez, and
  Pach\'{o}n]{DGP10}
T.~Dittrich, E.~A. G\'{o}mez and L.~A. Pach\'{o}n, \emph{J. Chem. Phys.}, 2010,
  \textbf{132}, 214102\relax
\mciteBstWouldAddEndPuncttrue
\mciteSetBstMidEndSepPunct{\mcitedefaultmidpunct}
{\mcitedefaultendpunct}{\mcitedefaultseppunct}\relax
\EndOfBibitem
\bibitem[Pach\'on \emph{et~al.}(2010)Pach\'on, Ingold, and Dittrich]{PID10}
L.~A. Pach\'on, G.-L. Ingold and T.~Dittrich, \emph{Chem. Phys.}, 2010,
  \textbf{375}, 209 -- 215\relax
\mciteBstWouldAddEndPuncttrue
\mciteSetBstMidEndSepPunct{\mcitedefaultmidpunct}
{\mcitedefaultendpunct}{\mcitedefaultseppunct}\relax
\EndOfBibitem
\bibitem[Makri and Makarov(1995)]{MD95}
N.~Makri and D.~E. Makarov, \emph{J. Chem. Phys.}, 1995, \textbf{102},
  4600--4610\relax
\mciteBstWouldAddEndPuncttrue
\mciteSetBstMidEndSepPunct{\mcitedefaultmidpunct}
{\mcitedefaultendpunct}{\mcitedefaultseppunct}\relax
\EndOfBibitem
\bibitem[Makri and Makarov(1995)]{MD95b}
N.~Makri and D.~E. Makarov, \emph{J. Chem. Phys.}, 1995, \textbf{102},
  4611--4618\relax
\mciteBstWouldAddEndPuncttrue
\mciteSetBstMidEndSepPunct{\mcitedefaultmidpunct}
{\mcitedefaultendpunct}{\mcitedefaultseppunct}\relax
\EndOfBibitem
\bibitem[Thorwart(2000)]{Tho00}
M.~Thorwart, \emph{Tunneling and vibrational relaxation in driven multilevel
  systems in driven multilevel systems}, Shaker Verlag, 2000\relax
\mciteBstWouldAddEndPuncttrue
\mciteSetBstMidEndSepPunct{\mcitedefaultmidpunct}
{\mcitedefaultendpunct}{\mcitedefaultseppunct}\relax
\EndOfBibitem
\bibitem[Thorwart \emph{et~al.}(2009)Thorwart, Eckel, Reina, Nalbach, and
  Weiss]{TE&09}
M.~Thorwart, J.~Eckel, J.~Reina, P.~Nalbach and S.~Weiss, \emph{Chem. Phys.
  Lett.}, 2009, \textbf{478}, 234 -- 237\relax
\mciteBstWouldAddEndPuncttrue
\mciteSetBstMidEndSepPunct{\mcitedefaultmidpunct}
{\mcitedefaultendpunct}{\mcitedefaultseppunct}\relax
\EndOfBibitem
\bibitem[Nalbach \emph{et~al.}(2011)Nalbach, Ishizaki, Fleming, and
  Thorwart]{NI&11}
P.~Nalbach, A.~Ishizaki, G.~R. Fleming and M.~Thorwart, \emph{New J. Phys.},
  2011, \textbf{13}, 063040\relax
\mciteBstWouldAddEndPuncttrue
\mciteSetBstMidEndSepPunct{\mcitedefaultmidpunct}
{\mcitedefaultendpunct}{\mcitedefaultseppunct}\relax
\EndOfBibitem
\bibitem[Hanson \emph{et~al.}(2007)Hanson, Kouwenhoven, Petta, Tarucha, and
  Vandersypen]{HK&07}
R.~Hanson, L.~P. Kouwenhoven, J.~R. Petta, S.~Tarucha and L.~M.~K. Vandersypen,
  \emph{Rev. Mod. Phys.}, 2007, \textbf{79}, 1217--1265\relax
\mciteBstWouldAddEndPuncttrue
\mciteSetBstMidEndSepPunct{\mcitedefaultmidpunct}
{\mcitedefaultendpunct}{\mcitedefaultseppunct}\relax
\EndOfBibitem
\bibitem[Grabert \emph{et~al.}(1984)Grabert, Weiss, and Talkner]{GWT84}
H.~Grabert, U.~Weiss and P.~Talkner, \emph{Z. Phys. B}, 1984, \textbf{55},
  87\relax
\mciteBstWouldAddEndPuncttrue
\mciteSetBstMidEndSepPunct{\mcitedefaultmidpunct}
{\mcitedefaultendpunct}{\mcitedefaultseppunct}\relax
\EndOfBibitem
\bibitem[Haake and Reibold(1985)]{HR85}
F.~Haake and R.~Reibold, \emph{Phys. Rev. A}, 1985, \textbf{32},
  2462--2475\relax
\mciteBstWouldAddEndPuncttrue
\mciteSetBstMidEndSepPunct{\mcitedefaultmidpunct}
{\mcitedefaultendpunct}{\mcitedefaultseppunct}\relax
\EndOfBibitem
\bibitem[Wu \emph{et~al.}(2010)Wu, Liu, Shen, Cao, and Silbey]{WL&10}
J.~Wu, F.~Liu, Y.~Shen, J.~Cao and R.~J. Silbey, \emph{New J. Phys.}, 2010,
  \textbf{12}, 105012\relax
\mciteBstWouldAddEndPuncttrue
\mciteSetBstMidEndSepPunct{\mcitedefaultmidpunct}
{\mcitedefaultendpunct}{\mcitedefaultseppunct}\relax
\EndOfBibitem
\bibitem[Wu \emph{et~al.}(2011)Wu, Ma, Wang, Silbey, and Cao]{WL&11}
F.~Wu, J.and~Liu, J.~Ma, X.~Wang, R.~Silbey and J.~Cao, 2011,
  arXiv:1109.5769\relax
\mciteBstWouldAddEndPuncttrue
\mciteSetBstMidEndSepPunct{\mcitedefaultmidpunct}
{\mcitedefaultendpunct}{\mcitedefaultseppunct}\relax
\EndOfBibitem
\bibitem[Haken and Strobl(1973)]{HS73}
H.~Haken and G.~Strobl, \emph{Z. Phys. A Hadron Nucl.}, 1973, \textbf{262},
  135--148\relax
\mciteBstWouldAddEndPuncttrue
\mciteSetBstMidEndSepPunct{\mcitedefaultmidpunct}
{\mcitedefaultendpunct}{\mcitedefaultseppunct}\relax
\EndOfBibitem
\bibitem[Cao(1997)]{Cao97}
J.~Cao, \emph{J. Chem. Phys.}, 1997, \textbf{107}, 3204--3209\relax
\mciteBstWouldAddEndPuncttrue
\mciteSetBstMidEndSepPunct{\mcitedefaultmidpunct}
{\mcitedefaultendpunct}{\mcitedefaultseppunct}\relax
\EndOfBibitem
\bibitem[Dekker(1987)]{Dek87}
H.~Dekker, \emph{Phys. Rev. A}, 1987, \textbf{35}, 1436--1437\relax
\mciteBstWouldAddEndPuncttrue
\mciteSetBstMidEndSepPunct{\mcitedefaultmidpunct}
{\mcitedefaultendpunct}{\mcitedefaultseppunct}\relax
\EndOfBibitem
\bibitem[Aslangul \emph{et~al.}(1986)Aslangul, Pottier, and Saint-James]{APS86}
C.~Aslangul, N.~Pottier and D.~Saint-James, \emph{J. Phys. France}, 1986,
  \textbf{47}, 1657--1661\relax
\mciteBstWouldAddEndPuncttrue
\mciteSetBstMidEndSepPunct{\mcitedefaultmidpunct}
{\mcitedefaultendpunct}{\mcitedefaultseppunct}\relax
\EndOfBibitem
\bibitem[Eckel \emph{et~al.}(2009)Eckel, Reina, and Thorwart]{ERT09}
J.~Eckel, J.~H. Reina and M.~Thorwart, \emph{New J. Phys.}, 2009, \textbf{11},
  085001\relax
\mciteBstWouldAddEndPuncttrue
\mciteSetBstMidEndSepPunct{\mcitedefaultmidpunct}
{\mcitedefaultendpunct}{\mcitedefaultseppunct}\relax
\EndOfBibitem
\bibitem[Scholes \emph{et~al.}(2007)Scholes, Curutchet, Mennucci, Cammi, and
  Tomasi]{SC&07}
G.~D. Scholes, C.~Curutchet, B.~Mennucci, R.~Cammi and J.~Tomasi, \emph{J.
  Phys. Chem. B}, 2007, \textbf{111}, 6978--6982\relax
\mciteBstWouldAddEndPuncttrue
\mciteSetBstMidEndSepPunct{\mcitedefaultmidpunct}
{\mcitedefaultendpunct}{\mcitedefaultseppunct}\relax
\EndOfBibitem
\bibitem[Jing \emph{et~al.}(2012)Jing, Zheng, Li, and Shi]{JZ&12}
Y.~Jing, R.~Zheng, H.-X. Li and Q.~Shi, \emph{The Journal of Physical Chemistry
  B}, 2012, \textbf{116}, 1164--1171\relax
\mciteBstWouldAddEndPuncttrue
\mciteSetBstMidEndSepPunct{\mcitedefaultmidpunct}
{\mcitedefaultendpunct}{\mcitedefaultseppunct}\relax
\EndOfBibitem
\bibitem[Tomasi \emph{et~al.}(2005)Tomasi, Mennucci, and Cammi]{TMC05}
J.~Tomasi, B.~Mennucci and R.~Cammi, \emph{Chem. Rev.}, 2005, \textbf{105},
  2999--3094\relax
\mciteBstWouldAddEndPuncttrue
\mciteSetBstMidEndSepPunct{\mcitedefaultmidpunct}
{\mcitedefaultendpunct}{\mcitedefaultseppunct}\relax
\EndOfBibitem
\bibitem[Curutchet \emph{et~al.}(2011)Curutchet, Kongsted, Mu\~noz Losa,
  Hossein-Nejad, Scholes, and Mennucci]{CK&11}
C.~Curutchet, J.~Kongsted, A.~Mu\~noz Losa, H.~Hossein-Nejad, G.~D. Scholes and
  B.~Mennucci, \emph{J. Am. Chem. Soc.}, 2011, \textbf{133}, 3078--3084\relax
\mciteBstWouldAddEndPuncttrue
\mciteSetBstMidEndSepPunct{\mcitedefaultmidpunct}
{\mcitedefaultendpunct}{\mcitedefaultseppunct}\relax
\EndOfBibitem
\bibitem[Damjanovi\ifmmode~\acute{c}\else \'{c}\fi{}
  \emph{et~al.}(2002)Damjanovi\ifmmode~\acute{c}\else \'{c}\fi{}, Kosztin,
  Kleinekath\"ofer, and Schulten]{DK&02}
A.~Damjanovi\ifmmode~\acute{c}\else \'{c}\fi{}, I.~Kosztin, U.~Kleinekath\"ofer
  and K.~Schulten, \emph{Phys. Rev. E}, 2002, \textbf{65}, 031919\relax
\mciteBstWouldAddEndPuncttrue
\mciteSetBstMidEndSepPunct{\mcitedefaultmidpunct}
{\mcitedefaultendpunct}{\mcitedefaultseppunct}\relax
\EndOfBibitem
\bibitem[Curutchet \emph{et~al.}(2009)Curutchet, Mu\~noz Losa, Monti, Kongsted,
  Scholes, and Mennucci]{CM&09}
C.~Curutchet, A.~Mu\~noz Losa, S.~Monti, J.~Kongsted, G.~D. Scholes and
  B.~Mennucci, \emph{J. Chem. Theory Comput.}, 2009, \textbf{5},
  1838--1848\relax
\mciteBstWouldAddEndPuncttrue
\mciteSetBstMidEndSepPunct{\mcitedefaultmidpunct}
{\mcitedefaultendpunct}{\mcitedefaultseppunct}\relax
\EndOfBibitem
\bibitem[Olbrich and Kleinekath\"ofer(2010)]{OK10}
C.~Olbrich and U.~Kleinekath\"ofer, \emph{J. Phys. Chem. B}, 2010,
  \textbf{114}, 12427--12437\relax
\mciteBstWouldAddEndPuncttrue
\mciteSetBstMidEndSepPunct{\mcitedefaultmidpunct}
{\mcitedefaultendpunct}{\mcitedefaultseppunct}\relax
\EndOfBibitem
\bibitem[Olbrich \emph{et~al.}(2011)Olbrich, Str\"umpfer, Schulten, and
  Kleinekath\"ofer]{OS&11}
C.~Olbrich, J.~Str\"umpfer, K.~Schulten and U.~Kleinekath\"ofer, \emph{J. Phys.
  Chem. B}, 2011, \textbf{115}, 758--764\relax
\mciteBstWouldAddEndPuncttrue
\mciteSetBstMidEndSepPunct{\mcitedefaultmidpunct}
{\mcitedefaultendpunct}{\mcitedefaultseppunct}\relax
\EndOfBibitem
\bibitem[M\"uh \emph{et~al.}(2007)M\"uh, Madjet, Adolphs, Abdurahman,
  Rabenstein, Ishikita, Knapp, and Renger]{MM&07}
F.~M\"uh, M.~E.-A. Madjet, J.~Adolphs, A.~Abdurahman, B.~Rabenstein,
  H.~Ishikita, E.-W. Knapp and T.~Renger, \emph{Proc. Natl. Acad. Sci. USA},
  2007, \textbf{104}, 16862--16867\relax
\mciteBstWouldAddEndPuncttrue
\mciteSetBstMidEndSepPunct{\mcitedefaultmidpunct}
{\mcitedefaultendpunct}{\mcitedefaultseppunct}\relax
\EndOfBibitem
\bibitem[Yu \emph{et~al.}(1999)Yu, Di\'osi, Gisin, and Strunz]{YDG&99}
T.~Yu, L.~Di\'osi, N.~Gisin and W.~T. Strunz, \emph{Phys. Rev. A}, 1999,
  \textbf{60}, 91--103\relax
\mciteBstWouldAddEndPuncttrue
\mciteSetBstMidEndSepPunct{\mcitedefaultmidpunct}
{\mcitedefaultendpunct}{\mcitedefaultseppunct}\relax
\EndOfBibitem
\bibitem[Stockburger and Grabert(2002)]{SG02}
J.~T. Stockburger and H.~Grabert, \emph{Phys. Rev. Lett.}, 2002, \textbf{88},
  170407\relax
\mciteBstWouldAddEndPuncttrue
\mciteSetBstMidEndSepPunct{\mcitedefaultmidpunct}
{\mcitedefaultendpunct}{\mcitedefaultseppunct}\relax
\EndOfBibitem
\bibitem[Ritschel \emph{et~al.}(2011)Ritschel, Roden, Strunz, Aspuru-Guzik, and
  Eisfeld]{RR&11}
G.~Ritschel, J.~Roden, W.~T. Strunz, A.~Aspuru-Guzik and A.~Eisfeld, \emph{J.
  Phys. Chem. Lett.}, 2011, \textbf{2}, 2912--2917\relax
\mciteBstWouldAddEndPuncttrue
\mciteSetBstMidEndSepPunct{\mcitedefaultmidpunct}
{\mcitedefaultendpunct}{\mcitedefaultseppunct}\relax
\EndOfBibitem
\bibitem[Ritschel \emph{et~al.}(2011)Ritschel, Roden, Strunz, and
  Eisfeld]{RR&11b}
G.~Ritschel, J.~Roden, W.~T. Strunz and A.~Eisfeld, \emph{New J. Phys.}, 2011,
  \textbf{13}, 113034\relax
\mciteBstWouldAddEndPuncttrue
\mciteSetBstMidEndSepPunct{\mcitedefaultmidpunct}
{\mcitedefaultendpunct}{\mcitedefaultseppunct}\relax
\EndOfBibitem
\bibitem[Moix \emph{et~al.}(2012)Moix, Zhao, and Cao]{MZC12}
J.~M. Moix, Y.~Zhao and J.~Cao, \emph{arXiv:1202.4705v1}, 2012\relax
\mciteBstWouldAddEndPuncttrue
\mciteSetBstMidEndSepPunct{\mcitedefaultmidpunct}
{\mcitedefaultendpunct}{\mcitedefaultseppunct}\relax
\EndOfBibitem
\bibitem[Olbrich and Kleinekath\"ofer(2010)]{OS&10}
C.~Olbrich and U.~Kleinekath\"ofer, \emph{J. Phys. Chem. B}, 2010,
  \textbf{114}, 12427--12437\relax
\mciteBstWouldAddEndPuncttrue
\mciteSetBstMidEndSepPunct{\mcitedefaultmidpunct}
{\mcitedefaultendpunct}{\mcitedefaultseppunct}\relax
\EndOfBibitem
\bibitem[Pengfei and Coker(2012)]{HC12}
H.~Pengfei and D.~F. Coker, \emph{J. Chem. Phys.}, 2012, \textbf{130}, \relax
\mciteBstWouldAddEndPuncttrue
\mciteSetBstMidEndSepPunct{\mcitedefaultmidpunct}
{\mcitedefaultendpunct}{\mcitedefaultseppunct}\relax
\EndOfBibitem
\bibitem[Christensson \emph{et~al.}(2012)Christensson, Kauffmann, Pullerits,
  and Mancal]{CK&12}
N.~Christensson, H.~F. Kauffmann, T.~Pullerits and T.~Mancal,
  \emph{arXiv:1201.6325v1}, 2012\relax
\mciteBstWouldAddEndPuncttrue
\mciteSetBstMidEndSepPunct{\mcitedefaultmidpunct}
{\mcitedefaultendpunct}{\mcitedefaultseppunct}\relax
\EndOfBibitem
\bibitem[Turner \emph{et~al.}(2012)Turner, Dinshaw, Lee, Belsley, Wilk, Curmi,
  and Scholes]{TD&12}
D.~B. Turner, R.~Dinshaw, K.-k. Lee, M.~Belsley, K.~E. Wilk, P.~M.~G. Curmi and
  G.~Scholes, \emph{Phys. Chem. Chem. Phys.}, 2012,  --\relax
\mciteBstWouldAddEndPuncttrue
\mciteSetBstMidEndSepPunct{\mcitedefaultmidpunct}
{\mcitedefaultendpunct}{\mcitedefaultseppunct}\relax
\EndOfBibitem
\bibitem[Elran and Brumer(2004)]{EB04}
Y.~Elran and P.~Brumer, \emph{J. Chem. Phys.}, 2004, \textbf{121},
  2673--2684\relax
\mciteBstWouldAddEndPuncttrue
\mciteSetBstMidEndSepPunct{\mcitedefaultmidpunct}
{\mcitedefaultendpunct}{\mcitedefaultseppunct}\relax
\EndOfBibitem
\bibitem[Hwang and Rossky(2004)]{HR04}
H.~Hwang and P.~J. Rossky, \emph{J. Chem. Phys.}, 2004, \textbf{120},
  11380\relax
\mciteBstWouldAddEndPuncttrue
\mciteSetBstMidEndSepPunct{\mcitedefaultmidpunct}
{\mcitedefaultendpunct}{\mcitedefaultseppunct}\relax
\EndOfBibitem
\bibitem[Gilmore and McKenzie(2008)]{GM08}
J.~B. Gilmore and R.~H. McKenzie, \emph{J. Phys. Chem. A}, 2008, \textbf{112},
  2162--2176\relax
\mciteBstWouldAddEndPuncttrue
\mciteSetBstMidEndSepPunct{\mcitedefaultmidpunct}
{\mcitedefaultendpunct}{\mcitedefaultseppunct}\relax
\EndOfBibitem
\bibitem[Briggs and Eisfeld(2011)]{BE11}
J.~S. Briggs and A.~Eisfeld, \emph{Phys. Rev. E}, 2011, \textbf{83},
  051911\relax
\mciteBstWouldAddEndPuncttrue
\mciteSetBstMidEndSepPunct{\mcitedefaultmidpunct}
{\mcitedefaultendpunct}{\mcitedefaultseppunct}\relax
\EndOfBibitem
\bibitem[Weiss and Wollensak(1989)]{WW89}
U.~Weiss and M.~Wollensak, \emph{Phys. Rev. Lett.}, 1989, \textbf{62},
  1663\relax
\mciteBstWouldAddEndPuncttrue
\mciteSetBstMidEndSepPunct{\mcitedefaultmidpunct}
{\mcitedefaultendpunct}{\mcitedefaultseppunct}\relax
\EndOfBibitem
\bibitem[Thoss \emph{et~al.}(2001)Thoss, Wang, and Miller]{TWM01}
M.~Thoss, H.~Wang and W.~H. Miller, \emph{J. Chem. Phys.}, 2001, \textbf{115},
  2991\relax
\mciteBstWouldAddEndPuncttrue
\mciteSetBstMidEndSepPunct{\mcitedefaultmidpunct}
{\mcitedefaultendpunct}{\mcitedefaultseppunct}\relax
\EndOfBibitem
\bibitem[Petta \emph{et~al.}(2005)Petta, Johnson, Taylor, Laird, Yacoby, Lukin,
  Marcus, Hanson, and Gossard]{PJ&05}
J.~R. Petta, A.~C. Johnson, J.~M. Taylor, E.~A. Laird, A.~Yacoby, M.~D. Lukin,
  C.~M. Marcus, M.~P. Hanson and A.~C. Gossard, \emph{Science}, 2005,
  \textbf{309}, 2180--2184\relax
\mciteBstWouldAddEndPuncttrue
\mciteSetBstMidEndSepPunct{\mcitedefaultmidpunct}
{\mcitedefaultendpunct}{\mcitedefaultseppunct}\relax
\EndOfBibitem
\bibitem[Habenicht \emph{et~al.}(2007)Habenicht, Kamisaka, Yamashita, and
  Prezhdo]{HK&07b}
B.~F. Habenicht, H.~Kamisaka, K.~Yamashita and O.~V. Prezhdo, \emph{Nano
  Lett.}, 2007, \textbf{7}, 3260--3265\relax
\mciteBstWouldAddEndPuncttrue
\mciteSetBstMidEndSepPunct{\mcitedefaultmidpunct}
{\mcitedefaultendpunct}{\mcitedefaultseppunct}\relax
\EndOfBibitem
\bibitem[Hoyer \emph{et~al.}(2010)Hoyer, Sarovar, and Whaley]{HSW10}
S.~Hoyer, M.~Sarovar and K.~B. Whaley, \emph{New J. Phys.}, 2010, \textbf{12},
  065041\relax
\mciteBstWouldAddEndPuncttrue
\mciteSetBstMidEndSepPunct{\mcitedefaultmidpunct}
{\mcitedefaultendpunct}{\mcitedefaultseppunct}\relax
\EndOfBibitem
\bibitem[Blankenship \emph{et~al.}(2011)Blankenship, Tiede, Barber, Brudvig,
  Fleming, Ghirardi, Gunner, Junge, Kramer, Melis, Moore, Moser, Nocera, Nozik,
  Ort, Parson, Prince, and Sayre]{BT&11}
R.~E. Blankenship, D.~M. Tiede, J.~Barber, G.~W. Brudvig, G.~Fleming,
  M.~Ghirardi, M.~R. Gunner, W.~Junge, D.~M. Kramer, A.~Melis, T.~A. Moore,
  C.~C. Moser, D.~G. Nocera, A.~J. Nozik, D.~R. Ort, W.~W. Parson, R.~C. Prince
  and R.~T. Sayre, \emph{Science}, 2011, \textbf{332}, 805--809\relax
\mciteBstWouldAddEndPuncttrue
\mciteSetBstMidEndSepPunct{\mcitedefaultmidpunct}
{\mcitedefaultendpunct}{\mcitedefaultseppunct}\relax
\EndOfBibitem
\bibitem[Hoyer \emph{et~al.}(2011)Hoyer, Ishizaki, and Whaley]{HIW11}
S.~Hoyer, A.~Ishizaki and K.~B. Whaley, 2011,  arXiv:1106.2911v1\relax
\mciteBstWouldAddEndPuncttrue
\mciteSetBstMidEndSepPunct{\mcitedefaultmidpunct}
{\mcitedefaultendpunct}{\mcitedefaultseppunct}\relax
\EndOfBibitem
\bibitem[Galve \emph{et~al.}(2010)Galve, Pach\'on, and Zueco]{GPZ10}
F.~Galve, L.~A. Pach\'on and D.~Zueco, \emph{Phys. Rev. Lett.}, 2010,
  \textbf{105}, 180501\relax
\mciteBstWouldAddEndPuncttrue
\mciteSetBstMidEndSepPunct{\mcitedefaultmidpunct}
{\mcitedefaultendpunct}{\mcitedefaultseppunct}\relax
\EndOfBibitem
\bibitem[Vedral(2010)]{Ved10}
V.~Vedral, \emph{Nature}, 2010, \textbf{468}, 769--770\relax
\mciteBstWouldAddEndPuncttrue
\mciteSetBstMidEndSepPunct{\mcitedefaultmidpunct}
{\mcitedefaultendpunct}{\mcitedefaultseppunct}\relax
\EndOfBibitem
\bibitem[Mart\'in-Delgado \emph{et~al.}(2002)Mart\'in-Delgado,
  Rodriguez-Laguna, and Sierra]{MRS02}
M.~A. Mart\'in-Delgado, J.~Rodriguez-Laguna and G.~Sierra, \emph{Phys. Rev. B},
  2002, \textbf{65}, 155116\relax
\mciteBstWouldAddEndPuncttrue
\mciteSetBstMidEndSepPunct{\mcitedefaultmidpunct}
{\mcitedefaultendpunct}{\mcitedefaultseppunct}\relax
\EndOfBibitem
\bibitem[Prior \emph{et~al.}(2010)Prior, Chin, Huelga, and Plenio]{PC&10}
J.~Prior, A.~W. Chin, S.~F. Huelga and M.~B. Plenio, \emph{Phys. Rev. Lett.},
  2010, \textbf{105}, 050404\relax
\mciteBstWouldAddEndPuncttrue
\mciteSetBstMidEndSepPunct{\mcitedefaultmidpunct}
{\mcitedefaultendpunct}{\mcitedefaultseppunct}\relax
\EndOfBibitem
\bibitem[Chin \emph{et~al.}(2012)Chin, Prior, Rosenbach, Caycedo-Soler, Huelga,
  and Plenio]{CP&12}
A.~W. Chin, J.~Prior, R.~Rosenbach, F.~Caycedo-Soler, S.~F. Huelga and M.~B.
  Plenio, \emph{Vibrational structures and long-lasting electronic coherence},
  2012,  arXiv:1203.0776v1\relax
\mciteBstWouldAddEndPuncttrue
\mciteSetBstMidEndSepPunct{\mcitedefaultmidpunct}
{\mcitedefaultendpunct}{\mcitedefaultseppunct}\relax
\EndOfBibitem
\end{mcitethebibliography}
\bibliographystyle{rsc} 
}


\end{document}